\documentclass[12pt, draftclsnofoot, onecolumn]{IEEEtran}
 \usepackage{amsmath,amssymb}
 \usepackage{subfigure}
 \usepackage{graphicx,graphics,color,psfrag}
 \usepackage{cite,balance}
 \usepackage{caption}
 \allowdisplaybreaks
 \usepackage{algorithm}
 \usepackage{accents}
 \usepackage{amsthm}
 \usepackage{bm}
 \usepackage{algorithmic}
 \usepackage[english]{babel}
 \usepackage{multirow}
 \usepackage{enumerate}
 \usepackage{cases}
 \usepackage{stfloats}
 \usepackage{dsfont}
 \usepackage{color,soul}
 \usepackage{amsfonts}
 \usepackage{cite,graphicx,amsmath,amssymb}
 \usepackage{subfigure}
 \usepackage{fancyhdr}
 \usepackage{hhline}
 \usepackage{graphicx,graphics}
 \usepackage{array,color}
 \usepackage{amsmath}
 \usepackage{booktabs}

\newtheorem{lemma}{Lemma}

\newtheorem{proposition}{Proposition}

\newtheorem{remark}{\bf Remark}

\def\qed{$\Box$}

\def\proof{\noindent{\emph{Proof:} }}

\def
\endproof{\hspace*{\fill}~\qed
\par
\endtrivlist\unskip}
\def\endproof{\hspace*{\fill}~\qed\par\endtrivlist\vskip3pt}

\def\E{\mathsf{E}}

\def\phi{\varphi}

\def\rank{{\operatorname{rank}}}

\def\l{\left}
\def\r{\right}
\def\({\left(}
\def\){\right)}

\def\ba{{\mathbf{a}}}

\def\bd{{\mathbf{d}}}

\def\bx{{\mathbf{x}}}
\def\by{{\mathbf{y}}}

\def\b0{{\mathbf{0}}}

\def\bI{{\mathbf{I}}}

\newcommand{\tr}{\mathrm{tr}}

\begin{document}
\setlength{\topskip}{-3pt}

\title{\huge Wirelessly Powered Data Aggregation for IoT via Over-the-Air Functional Computation: Beamforming and Power Control}
\author{Xiaoyang Li, Guangxu Zhu, Yi Gong, and Kaibin Huang
\thanks{X. Li, G. Zhu, and K. Huang are with the Dept. of Electrical \& Electronic Engr.  (EEE) at The University of Hong Kong, Hong Kong (e-mail: lixy@eee.hku.hk, gxzhu@eee.hku.hk; huangkb@eee.hku.hk). X. Li is also with the Dept. of EEE at Southern University of Science and Technology, Shenzhen, China. Y. Gong is affiliated with the same department (e-mail: gongy@sustc.edu.cn). Corresponding author: K. Huang.} 
}
\maketitle

\vspace{-18mm}
\begin{abstract}
As a revolution in networking, \emph{Internet of Things} (IoT) aims at automating the operations of our societies by  connecting and leveraging an enormous number of distributed devices (e.g., sensors and actuators). One design challenge is efficient \emph{wireless data aggregation} (WDA) over tremendous IoT devices. This can enable a series of IoT applications ranging from latency-sensitive high-mobility sensing to data-intensive distributed machine learning. \emph{Over-the-air (functional) computation} (AirComp) has emerged to be a promising solution that merges computing and communication by exploiting analog-wave addition in the air. Another IoT design challenge is battery recharging for dense sensors which can be tackled by \emph{wireless power transfer} (WPT). The coexisting of AirComp and WPT in IoT system calls for their integration to enhance the performance and efficiency of WDA. This motivates the current work on developing the \emph{wirelessly powered AirComp} (WP-AirComp) framework by jointly optimizing wireless  power control, energy and (data) aggregation beamforming to minimize the AirComp error. To derive a practical solution, we recast the non-convex joint optimization problem into the equivalent \emph{outer and inner sub-problems} for (inner) wireless power control and energy beamforming, and (outer) the efficient aggregation beamforming, respectively. The former is solved in closed form while the latter is efficiently solved using the semidefinite relaxation technique. The results reveal that the optimal energy beams point to the dominant eigen-directions of the WPT channels, and the optimal power allocation tends to equalize the close-loop (down-link WPT and up-link AirComp) effective channels of different sensors. Simulation demonstrates that controlling WPT provides additional design dimensions for substantially reducing the AirComp error.
\end{abstract}

\vspace{-4mm}
\section{Introduction}
\vspace{-1mm}
In the near future, tens of billions  of \emph{Internet-of-things} (IoT) devices (e.g., sensors and actuators) are expected to be deployed to automate the operations of our societies and make the ambient environment smart. Among others, there exist two design challenges for IoT. The first is fast \emph{wireless data aggregation} (WDA), namely fast collection and processing of data distributed at tremendous IoT devices by wireless transmission. WDA is an enabling operation for a series of IoT applications such as fusion of sensing values in environmental monitoring \cite{giridhar2006toward}, aggregation of mobile updates in federated machine learning \cite{mcmahan2016communication}, and distributed consensus in fleet driving \cite{di2015distributed}. Fast WDA is needed to regulate latency in cases with ultra-dense devices and/or high mobility (e.g., for sensors carried by drones or vehicles). A promising solution is \emph{over-the-air (functional) computation} (AirComp), which realizes fast WDA by simultaneous transmissions and exploiting analog-wave addition in a multi-access channel \cite{KatabiAirComp2016}. The other design challenge for IoT is powering dense energy-constrained sensors for WDA and other operations. One attractive solution is \emph{wireless power transfer} (WPT) using microwaves, whose feasibility has been proven in practical sensor networks \cite{choi2017wireless}.

To facilitate efficient implementation for IoT, it is natural to tackle the two said challenges simultaneously by pursuing the fusion of two corresponding technologies: AirComp and WPT. The resultant design challenge lies in the joint optimization of their key operations at servers (or fusion centers), namely energy beamforming and power control for WPT， and (data) aggregation beamforming for AirComp. This motivates the current work on developing a framework called \emph{wirelessly powered over-the-air computation} (WP-AirComp).

\vspace{-4mm}
\subsection{WDA via Over-the-Air Computation}
\vspace{-1mm}
Data aggregation in IoT can be posed as the mathematical problem of computing at a server a function $h(\cdot)$ of distributed data samples, denoted as $\{\bx_k\}$, which is characterized by the following form:
\begin{align}\label{WDA_func}
\by=h(\bx_1, \bx_2, \cdots, \bx_K) = f\l(\sum_{k=1}^{K}g_{k}(\bx_k)\r),
\end{align}
where $f(\cdot)$ and $g_{k}(\cdot)$ represent post-processing at the server and pre-processing at a device, respectively. The class of functions having the above form is known as \emph{nomographic functions} such as averaging and geometric mean. Typical functions in this class are summarized in Table \ref{summary:table1}. The conventional approach for WDA decouples \emph{data collection}, namely distributed transmission of $\{\bx_k\}$, and \emph{functional computation}, namely computing $h(\{\bx_k\})$ in \eqref{WDA_func}. The approach is incapable of supporting fast WDA as the application of any traditional orthogonal multi-access scheme for data collection results in the linear scaling of latency with the number of devices. In contrast, AirComp merges data collection (or else radio resources)  and computation. Specifically, the summation in \eqref{WDA_func} is implemented by simultaneous analog transmission to exploit the wave-addition of the multi-access channel. Consequently, the functional computation is performed "over-the-air" and the result is directly received by the server, thus giving the name of the technology. Simultaneous transmission in AirComp achieves low latency independent of the number of devices, and thereby enables fast WDA.

\begin{table}[t]
\centering
\caption{ Examples of nomographic functions that are AirComputable.}
\begin{tabular}{|p{4cm}|p{4cm}|}
\hline
\bf{Name} & \bf{Expression} \\
\hline
Arithmetic Mean &  $y = \frac{1}{K}\sum_{k=1}^K x_k$ \\ 
\hline
Weighted Sum &  $y = \sum_{k=1}^K \omega_k x_k$ \\ 
\hline
Geometric Mean &  $y = \l(\prod_{k=1}^K x_k \r)^{1/K}$ \\ 
\hline
Polynomial &  $y = \sum_{k=1}^K \omega_k x_k^{\beta_k}$ \\ 
\hline
Euclidean Norm &  $y = \sqrt{\sum_{k=1}^K x_k^2}$ \\ 
\hline
\end{tabular}
\label{summary:table1}
\vspace{-5mm}
\end{table}

The idea of AirComp can be traced back to the pioneering work studying functional computation in sensor networks \cite{GastparTIT2007}. In \cite{GastparTIT2007},  structured codes (e.g., lattice codes) are designed for reliable functional computation at a server based on distributed sensing values analog modulated and transmitted over a 、\emph{multi-access channel}. The importance of the work lies in the counter-intuitive finding that interference caused by simultaneous transmission can be exploited to facilitate computation. Subsequently, it was proved that the simple analog transmission without coding is optimal in terms of minimizing functional distortion in the case of independent Gaussian data sources \cite{GastparTIT2008}. Nevertheless, coding is still useful if the sources follow more complex distributions, such as bivariate Gaussian \cite{wagner2008rate}, correlated Gaussian \cite{SoundararajanTIT2012}, and sum of independent Gaussian \cite{GastparTIT2007}. The promising performance (with optimality in certain cases) of simple \emph{analog AirComp} has led to an active area focusing on its robustness and implementation \cite{GoldsmithTSP2008, GoldenbaumWCL2014, zhu2018over, KatabiAirshare2015,KatabiAirComp2016,GoldenbaumTCOM2013}. In particular, techniques for distributed power control and robust AirComp against channel estimation errors are proposed in \cite{GoldsmithTSP2008} and \cite{GoldenbaumWCL2014}, respectively. Another vein of research focuses on transforming AirComp from theory into practice by prototyping \cite{KatabiAirComp2016} and addressing practical issues such as transmission synchronization over sensors \cite{GoldenbaumTCOM2013,KatabiAirshare2015}. 

It is also worth mentioning that inspired by the success of AirComp in computation-centric sensor networks, the relevant principles have been applied to design new schemes for rate-centric communication networks. The \emph{compute-and-forward} relaying schemes as proposed in \cite{GastparTIT2011} decodes and forwards linear functions of the transmitted messages. The \emph{integer-forcing} linear receiver designed in \cite{GastparTIT2014} spatial multiplexes lattice codewords. Furthermore, the well-known \emph{physical layer network coding} leveraging analog wave addition generalizes the celebrated network coding schemes invented for wired networks to wireless networks \cite{GastparProceedings2011}.

IoT will feature \emph{multi-modal sensing} and the prevalence of antenna arrays (especially highly compact millimeter-wave arrays) at both servers and devices \cite{gubbi2013internet}. A multi-modal sensor can simultaneous monitor multiple parameters of the environment (e.g., pressure, light, humidity, and pollution), or of a user state (e.g., motion, location, and speech). The need of WDA over multi-modal sensors and other data-intensive IoT applications (e.g., federated machine learning) calls for the acceleration of computation rates in AirComp. While prior works mostly target single-antenna sensor network and support scalar-function AirComp, recent research has started to explore the possibility of using antenna arrays to enable the vector-function AirComp. A technique called \emph{multiple-input-multiple-output} (MIMO) AirComp leverages spatial \emph{degrees-of-freedom} to spatially multiplex multi-function computation (i.e., vector-function computation) and reduce computation errors by noise suppression \cite{zhu2018over}. The key design challenge lies in the optimization of the receive beamformer, called \emph{aggregation beamformer}, with the objective of minimizing the error of computed functions, w.r.t., the desired result in \eqref{WDA_func}. On the other hand, traditional multi-user MIMO beamforming aims at a different objective that is to minimize the sum distortion of individual data symbols. The difference in objective results in a fundamental shift in the beamforming design principle. To be specific, multi-user beamforming attempts to decouple simultaneously transmitted symbols (or equivalently sum-rate maximization), yielding the classic zero-forcing or \emph{minimizing mean-squared-error} (MMSE) design \cite{klein1996zero}. In contrast, aggregation beamforming tries to balance the effects of individual MIMO channels so as to accurately compute the function in \eqref{WDA_func}. The principle is reflected in the beamformer design in \cite{zhu2018over} that is obtained from the weighted sum of multi-user MIMO channels projected onto a Grassmann manifold.

\vspace{-3mm}
\subsection{Energy Beamforming for Wirelessly Powered Communication}
\vspace{-1mm}
WPT was originally developed for point-to-point power delivery using radio waves \cite{Zeng2016Communications}. Recently, the technology has been further developed to power communication networks  \cite{huang2014cutting}. One research focus is \emph{energy beamforming}, referring to the use of an antenna array at a transmitter to beam energy in the direction of a targeted receiver \cite{bi2015wireless,xia2015efficiency}. The integration of WPT with traditional wireless communication has created a new area, called \emph{simultaneous wireless information and power transfer} (SWIPT), which remains active. Recent studies focus on applying SWIPT to a series of communication systems, including MIMO communication \cite{park2013joint,yang2015throughput}, \emph{orthogonal frequency-division multiple access} (OFDMA) \cite{ng2013wireless,zhang2016energy}, two-way transmission \cite{popovski2013interactive}, relaying \cite{ding2014wireless,zhong2014wireless,zheng2017resource}, and cognitive networking~\cite{ng2016multiobjective}. Practical SWIPT designs accounting for imperfect channel state information were developed in \cite{liu2015simultaneous}.
More recently, WPT was also considered for \emph{unmanned aerial vehicle} (UAV) assisted communication \cite{zeng2016throughput}, where an UAV serves as a mobile relay and its transmit power and trajectory are jointly optimized to maximize the throughput. 

Another important application of WPT is in sensor networks, with energy constrained sensors. Recharging the batteries of hundreds to thousands of sensors is a challenging problem that can be solved efficiently by WPT \cite{choi2017wireless}. For multi-user WPT systems, energy beamforming at different power beacons needs to be coordinated to enable efficient energy multicasting. To this end, a collaborative energy-beamforming scheme is proposed in \cite{Xu2016Wireless} for efficiently powering a sensor network. For large-scale wirelessly powered sensor networks, a novel framework of backscatter sensing was recently proposed in \cite{zhu2018inference}, where low-cost passive sensors upload their sensing data to a drone mounted reader by concurrently reflecting the beamed power signal from a power beacon in a designed probabilistic manner. Then statistical inference algorithms can be devised for sensing value recovery without the knowledge of channel state information.

Though energy beamforming in wirelessly powered communication and sensor networks have been widely investigated, most of prior work focuses on rate maximization via optimizing the WPT efficiency. The design of wirelessly powered in-network computation, such as WP-AirComp in this work, remains as uncharted area.

\vspace{-3mm}
\subsection{Contributions and Organization}
\vspace{-1mm}



We consider an IoT system supporting down-link WPT and up-link AirComp. To be specific, a multi-antenna server transmits energy to power multiple sensors (or other types of devices) so that they can transmit sensing data for WDA at the server based on AirComp. The server controls the amount of energy harvested by sensors via energy beamforming and power allocation to different beams. Since the transmit power of sensors depends on the harvested energy, the two WPT operations affect the analog-signal superposition in the array observations at the server. This introduces \emph{coupling between WPT and AirComp}, and hence necessitates the joint design of wireless power control, energy and (data) aggregation beamforming, which is the core of WP-AirComp development. Via the joint design, WPT contributes additional dimensions for reducing the computation error in AirComp.
The contributions of this work are summarized as follows.

\begin{itemize}
\item {\bf Decomposition Based Design Approach}: The said joint design is formulated as an optimization problem for minimizing the computation error in AirComp under  a power constraint for WPT. Without compromising optimality, a decomposition approach is derived that decomposes the non-convex problem into two tractable nested sub-problems. The inner sub-problem concerns WPT optimization, involving the joint design of power control and energy beamforming, given the aggregation beamformer. The outer sub-problem is the optimization of aggregation beamformer. The approach is applied to develop an efficient convex-optimization algorithm for solving the non-convex joint design problem of the simplified \emph{multiple-input single-output} (MISO) case with single-antenna sensors, corresponding to \emph{scalar-function AirComp}. The results are then extended to the general MIMO case with multi-antenna sensors, corresponding to \emph{vector-function AirComp}.

\item {\bf Wireless Power Control and Energy Beamforming}: The said inner sub-problem is solved in closed form. The solution has simple structures that facilitate the implementation of WP-AirComp. First, the optimal policy for wireless power control attempts to equalize the multiple cascaded WPT-AirComp (or down-link-up-link) channels to facilitate the outer sub-problem of designing aggregation beamformer. Second, each optimal energy beam points to the direction of the corresponding WPT channel vector in the MISO case or the dominant eigenvector of the channel matrix in the MIMO case.

\item {\bf Data Aggregation Beamforming}: The said outer sub-problem is shown to be solvable by the powerful technique of \emph{semi-definite relaxation} (SDR). It is demonstrated by simulation that the joint design with WPT leads to significant gain in computation accuracy compared with decoupled designs.

\end{itemize}

\emph{Organization}: The remainder of the paper is organized as follows. Section II introduces the WP-AirComp system model. Section III presents the problem formulation for the joint design of power control, energy and aggregation beamforming. The solution for the MISO case is presented in Section IV. The extension to the MIMO case is given in Section V. Section VI further analyzes the difference between WP-AirComp framework and traditional designs. Simulation results are provided in Section VII, followed by concluding remarks in Section VIII.

\emph{Notation}: Boldface lowercase symbols denote vectors (e.g., $\bold{a}$) and boldface uppercase symbols denote matrices (e.g., $\bold{A}$). $\mathbb{R}$, $\mathbb{R}_{++}$ and $\mathbb{C}$ denote the real domain, positive real domain and complex domain, respectively. $\mathbb{C}^{M \times N}$ denotes the space of vector/matrix. $\bold{A}^H$ and $\bold{A}^{-1}$ represent the conjugate transpose and inverse operations on a matrix $\bold{A}$, respectively. $\text{tr}(\bold{A})$ and $\lambda_{\min} (\bold{A})$ represent the trace and minimum eigenvalue of the square matrix $\bold{A}$, respectively. $\mathsf{E}\{\cdot\}$ denotes the expectation of a vector/matrix, $\bI$ denotes the identity matrix, $\left\| \cdot \right\|$ denotes the Euclidean norm. Last, $\mathcal{CN}(\mu, \sigma_n^2)$ denotes the complex Gaussian distribution with mean $\mu$ and variance $\sigma_n^2$.

\begin{figure}[t]
\centering
\includegraphics[scale=0.43]{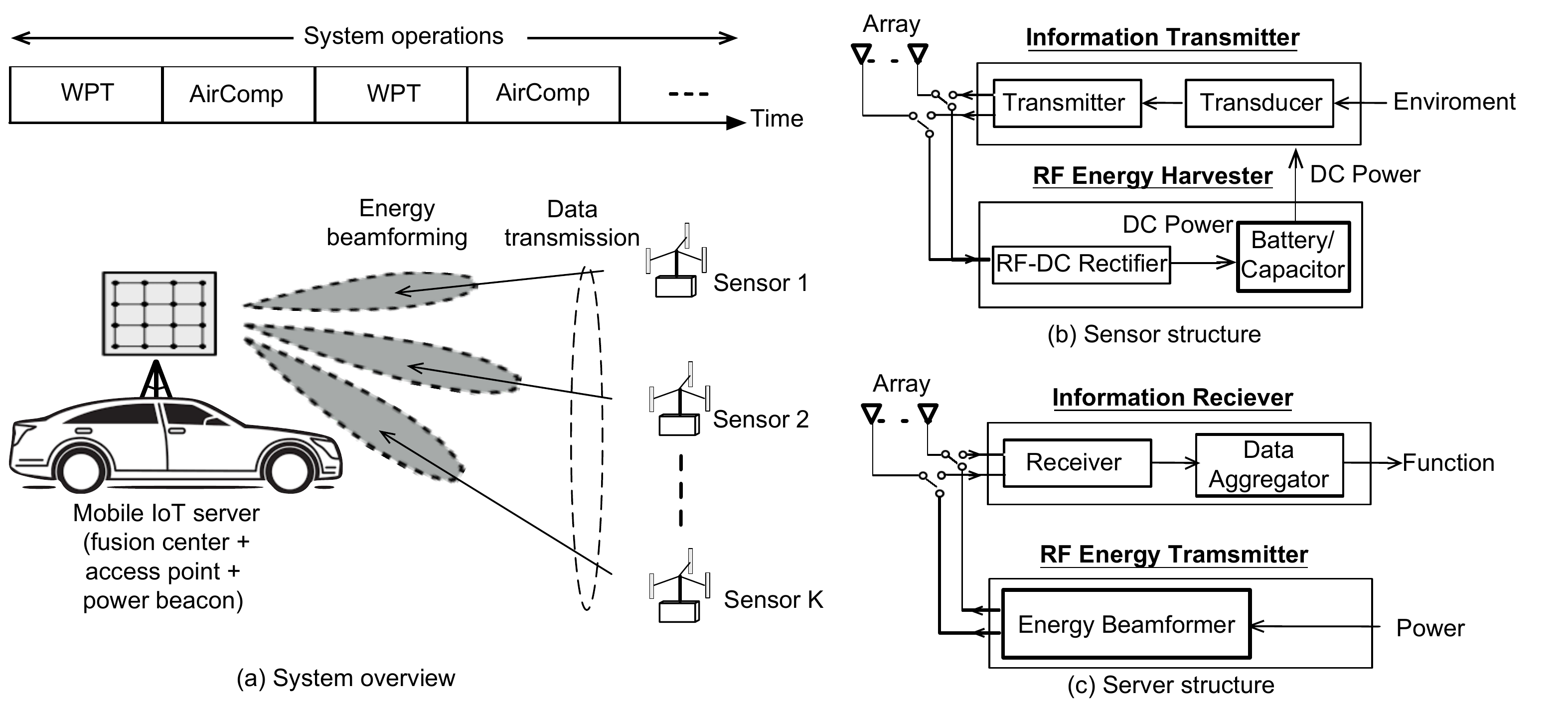}
\caption{A high-mobility IoT system aggregating distributed sensing data using WP-AirComp.}
\label{FigSys}
\end{figure}

\section{System Model}
We consider the mobile IoT system illustrated in Fig. \ref{FigSys}, where WP-AirComp is deployed for fast WDA. As shown in Fig \ref{FigSys} (c), the mobile server is multi-functional  serving as a power beacon, an access point and a data fusion center. The server is provisioned with an array of $N_{\sf AP}$ antennas. We consider both the cases of single-antenna and multi-antenna sensors, corresponding to the equivalent cases of MISO and MIMO channels. Let $N_{\sf SN}$ denote the number of antennas at each of total $K$ sensors. Time is divided into symbol durations, each of which lasts $t_0$ seconds and is called a \emph{(time) slot}. WP-AirComp is implemented based on the \emph{harvested-then-transmit} protocol that alternates WPT and AirComp phases with corresponding fixed durations. The energy a sensor harvests in a WPT phase is applied to power transmission in the following AirComp phase. The operations of different sensors are synchronized using a reference clock broadcast by the server (see e.g., \cite{KatabiAirshare2015}). For simplicity, channels are assumed to vary following the \emph{block-fading} model. In other words, each channel remains fixed within a (WPT/AirComp) phase and varies over different phases.

\begin{figure}[t]
\centering
\includegraphics[scale=0.43]{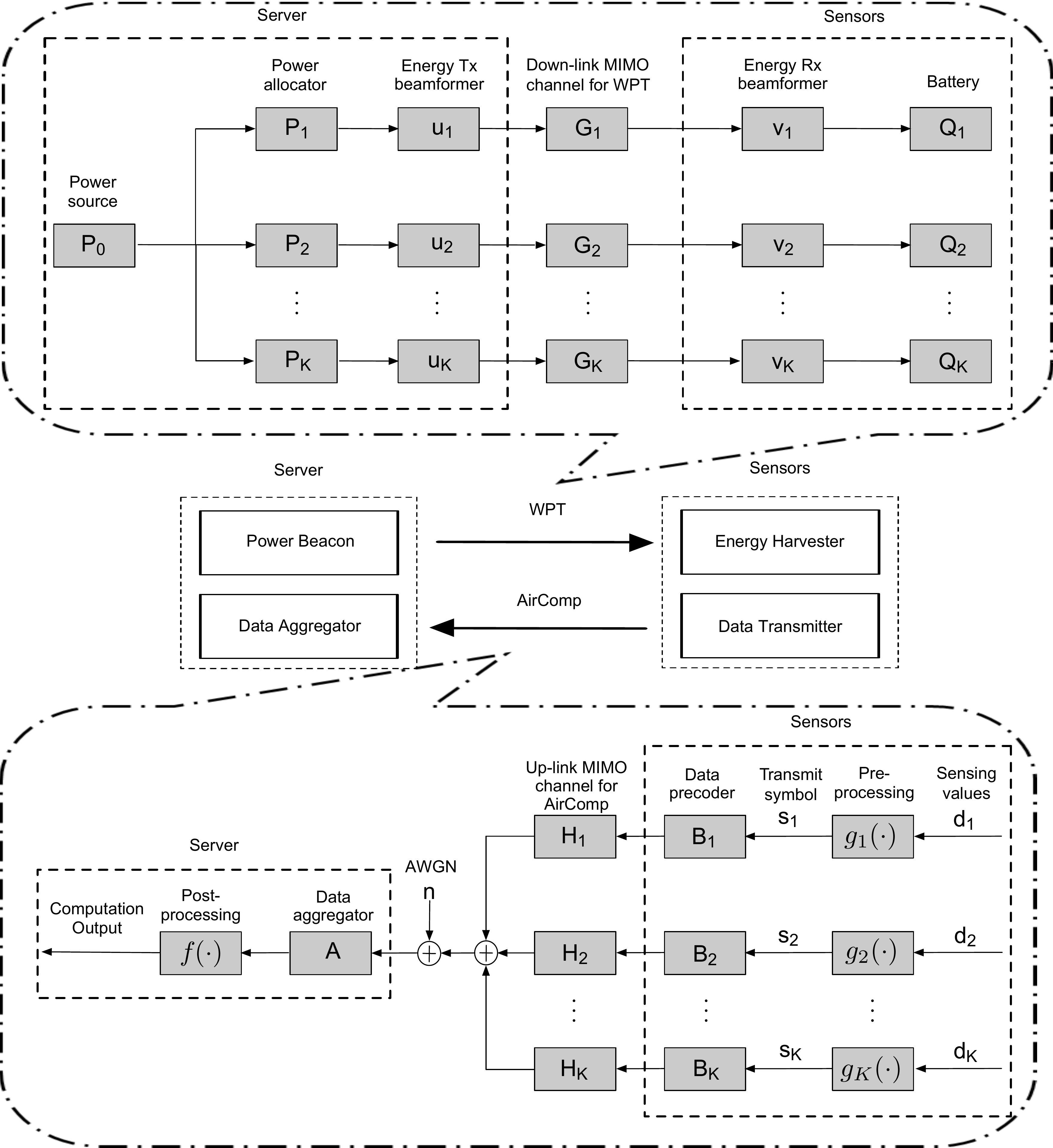}
\caption{Block diagram of WP-AirComp operations.}
\label{Mod_MIMO}
\end{figure}

\subsection{WPT Phase}
Consider an arbitrary WPT phase. The server serves as a power beacon that allocates power for different sensors and delivers the power by energy beamforming. Let $\bold{G}_k \in \mathbb{C}^{N_{\sf AP} \times N_{\sf SN}}$ represent the MIMO channel for the link from the server to sensor $k$, $P_k$ denote the power allocated to sensor $k$, and $\bold{u}_k\in\mathbb{C}^{N_{\sf AP}\times1}$ with $\bold{u}_k^H\bold{u}_k=1$ denote the transmit beamforming vector. The server transmission power is limited by a fixed value $P_0$, giving the following total power constraint:
\begin{equation} \label{Eq:Ptotal}
\sum_{k=1}^K P_k \le P_0.
\end{equation}
At sensor $k$, a receive beam is steered to harvested the transferred energy with the beamforming vector denoted as $\bold{v}_k\in\mathbb{C}^{N_{\sf SN}\times1}$ and $\bold{v}_k^H\bold{v}_k=1$.

Assuming energy beams are sufficiently sharp such that harvesting the side lobes of unintended beams gives a sensor negligible energy. Moreover, the energy-conversion process at a sensor, say sensor $k$, is characterized by a fixed efficiency factor, denoted as $\alpha_k$. Then the amount of energy harvested by sensor $k$ in a WPT phase of $T$ slots is given by
\begin{equation} \label{Eq:harvested}
Q_k = \alpha_k \|\bold{u}_k^H\bold{G}_k\bold{v}_k\|^2 P_k T.
\end{equation}

\subsection{AirComp Phase}
Consider an arbitrary AirComp phase. Each sensor records the values of $L$ heterogeneous parameters of an external time-varying process such as the ambient environment, or a human being. The measurement generates a vector symbol in each slot, represented by $\bold{d}_k=[d_{k1},d_{k2},...,d_{kL}]^T\in \mathbb{R}^{L \times 1}$ for sensor $k$. Powered by WPT, each sensor transmits vector symbols using its array to the server. On the other hand, the server doubly serves as an access point and a fusion center in the current phase. For WDA, the server aims at computing a \emph{vector function} of the distributed vector symbols $\{\bd_{k}\}_{k=1}^{K}$. Let $y = [y_1, y_2, \cdots, y_L ] \in \mathbb{R}^{L \times 1}$ denote the desired computation output, called the \emph{target-function vector}. Then $y_\ell$ is a nomographic function, denoted as $h_\ell$, of $K$ simultaneous observations of parameter $\ell$, namely $\{d_{k\ell}\}_{k=1}^{K}$. Following the definition in \eqref{WDA_func}, we have
\begin{align}\label{Nomographic}
y_{\ell} = h_\ell (d_{1\ell},d_{2\ell},\cdots,d_{K\ell})  = f_{\ell}\l(\sum_{k=1}^{K}g_{k\ell}(d_{k\ell})\r),
\end{align}
where $f_{\ell}(\cdot)$ and $\{g_{k\ell}(\cdot)\}$ represent the post-processing and pre-processing functions (see Table \ref{summary:table1} for examples). Let the vector symbol pre-processed and transmitted by sensor $k$ using linear analog modulation be denoted as $\bold{s}_k =[g_{k1}(d_{k1}), g_{k2}(d_{k2}), \cdots, g_{kL}(d_{kL})]^T$. To facilitate power control but without loss of generality, $\bold{s}_k$ is assumed to have unit variance, i.e., $\mathsf{E}\{\bold{s}_k \bold{s}_k^H\} = \bold{I}$. The WDA process and performance metric are described as follows.

Given synchronized symbol boundaries, all sensors transmit their vector symbols simultaneously. The distortion of array observations at the server with respect to the target-function vectors due to channel noise is suppressed using aggregation beamforming. To this end, the WDA process attempts to achieve coherent combining of $K$ vector symbols at the server in each slot. Let $\bold{A} \in \mathbb{C}^{N_{\sf AP} \times L}$ denote the aggregation beamforming matrix and $\bold{B}_k \in \mathbb{C}^{N_{\sf SN} \times L}$ the data precoding matrix at sensor $k$. Then the vector symbol received by the server after beamforming is given by: 
\begin{equation}\label{estimated_s}
\hat{\bold{s}} = \bold{A}^H \sum_{k=1}^K \bold{H}_k \bold{B}_k \bold{s}_k + \bold{A}^H \bold{n},
\end{equation}
where $\bold{H}_k \in \mathbb{C}^{N_{\sf AP} \times N_{\sf SN}}$ represents the MIMO channel for the link from sensor $k$ to the server, and $\bold{n}$ is the \emph{additive white Gaussian noise} (AWGN) vector with \emph{independent and identically distributed} (i.i.d.) $\mathcal{CN}(0, \sigma_n^2)$ elements. The distortion of $\hat{\bold{s}}$ with respect to the desired vector $\bold{s}$, called the \emph{computation error}, quantifies the AirComp performance and is measured by the \emph{mean-squared-error} (MSE) defined as
\begin{equation}\label{def:MSE}
 \textsf{MSE}(\hat{\bold{s}},\bold{s})=\E\l[(\hat{\bold{s}}-\bold{s})(\hat{\bold{s}}-\bold{s})^H\r].
\end{equation}
Substituting \eqref{estimated_s} into \eqref{def:MSE}, the MSE can be explicitly written as a function of the aggregation beamforming and precoding as follows:
\begin{equation}\label{MSE_function}
\textsf{MSE}(\bold{A}, \{\bold{B}_k\}) = \sum_{k=1}^{K} \tr((\bold{A}^H\bold{H}_k \bold{B}_k-\bold{B}_k)(\bold{A}^H\bold{H}_k \bold{B}_k-\bold{B}_k)^H)+\sigma_n^2\bold{A}^H\bold{A}.
\end{equation}
The aggregation beamforming and precoding are jointly optimized in the sequel under the criteria of MMSE. Since the transmission energy of each sensor cannot exceed the harvested energy, the transmission power constraint for sensor $k$ can be obtained using \eqref{Eq:harvested} as
\begin{equation}\label{Eq:Pprecoder}
\tr(\bold{B}_k\bold{B}^H_k) \le \frac{\alpha_k \theta_k T}{T_0} \|\bold{u}_k^H\bold{G}_k\bold{v}_k\|^2 P_k,
\end{equation}
where $T_0$ denotes the number of vector symbols transmitted within each AirComp phase and the factor $\theta_k \in (0,1)$ represents the fraction of harvested energy allocated for transmission. For ease of notation, we denote $\gamma_k = \frac{\alpha_k \theta_k T}{T_0}$ as the effective power conversion efficiency accounting for all the power discounting factors in the subsequent analysis.

\begin{figure}[t]
  \centering
  \includegraphics[scale=0.43]{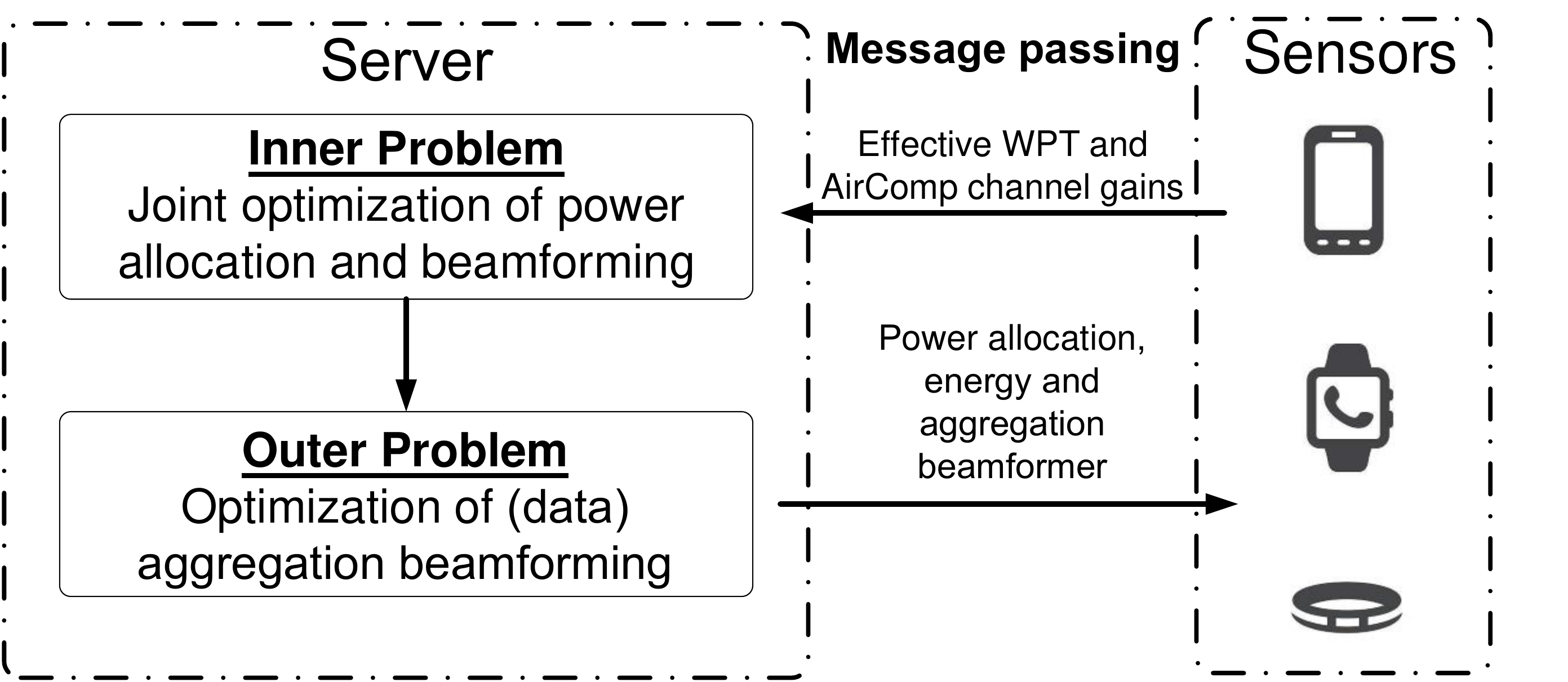}
  \caption{Outer-inner problem in WP-AirComp design and the required signaling procedure.}
  \label{FigOutin}
\end{figure}

\section{Joint Design of WPT and AirComp: Problem Formulation}
The joint design of the WPT and AirComp phases in the WP-AirComp system can be formulated as a joint optimization problem over wireless power control, energy and aggregation beamforming. Specifically, given the computation error in \eqref{MSE_function} and the two power constraints in \eqref{Eq:Ptotal} and \eqref{Eq:Pprecoder}, the problem can be formulated as:
\begin{equation*}\textbf{(P1)}\qquad
\begin{aligned}
\min_{\substack{ \bold{A}, \{\bold{B}_k\}, \\ \{\bold{u}_k\}, \{\bold{v}_k\},\{P_k\}}} \quad 
&\sum_{k=1}^{K}\tr((\bold{A}^H \bold{H}_k\bold{B}_k-\bold{I})(\bold{A}^H \bold{H}_k\bold{B}_k-\bold{I})^H) + \sigma_n^2\tr(\bold{A}^H\bold{A})\\ 
\text{s.t.} \quad 
&\tr(\bold{B}_k\bold{B}^H_k) \le \gamma_k \|\bold{u}_k^H\bold{G}_k\bold{v}_k\|^2 P_k, \forall k,\\
&\sum_{k=1}^{K} P_k \le P_0,\\
&\bold{u}_k^H\bold{u}_k=1,~\forall k,\\
&\bold{v}_k^H\bold{v}_k=1,~\forall k.
\end{aligned}
\end{equation*}

Problem P1 is difficult to solve due to its \emph{non-convexity}. The lack of convexity arises from the coupling between the transmit and receive beamformers for both the WPT and AirComp phases. To simplify the problem and shed light on the optimal solution structure, we first consider the simplified case of single-antenna sensors, resulting in MISO channels, and thus called the MISO case. Note that in the MISO case, only scalar-function AirComp is feasible. In the optimization problem for this case, the receive energy beamformers at sensors varnish and the data precoders reduce to scalars denoted as $\{b_k\}$, making the solution tractable. Let the $k$-th vector channels for AirComp and WPT be denoted as $\bold{h}_k \in \mathbb{C}^{N_{\sf AP}\times 1}$ and $\bold{g}_k\in\mathbb{C}^{N_{\sf AP}\times1}$ respectively, and the aggregation beamformer vector as $\bold{a} \in \mathbb{C}^{N_{\sf AP} \times 1}$. Then the original problem in P1 is simplified for the MISO case as:
\begin{equation*}\textbf{(P2)}\qquad
\begin{aligned}
\min_{\bold{a}, \{b_k\}, \{\bold{u}_k\}, \{P_k\}} \quad 
&\sum_{k=1}^{K} \|\bold{a}^H\bold{h}_k b_k-1\|^2+\sigma_n^2\bold{a}^H\bold{a}\\ 
\text{s.t.} \quad 
&\|b_k\|^2 \le \gamma_k \|\bold{u}_k^H\bold{g}_k\|^2 P_k,~\forall k,\\
&\sum_{k=1}^K P_k \le P_0,\\
&\bold{u}_k^H \bold{u}_k = 1,~\forall k.
\end{aligned}
\end{equation*}
Problem P2 is solved in the next section. The insights are leveraged to develop a practical solution in Section V. Last, we remark that an alternative and also natural formulation based on maximization of the receive \emph{signal-to-noise ratio} (SNR) leads to strategies fundamentally different from computation-error minimization. More details are given in Section VI.

\section{Joint Design of WPT and AirComp: MISO Case}
In this section, we consider the MISO case corresponding to single-antenna sensors and solve the design problem in P2. To this end, a decomposition based solution approach is developed. The approach and the solution of Problem P2 are discussed in the following sub-sections.

\subsection{Decomposition Approach}
Problem P2 remains non-convex and difficult to solve directly. We overcome the difficulty by decomposing it into two solvable sub-problems. To begin with, \emph{channel-inversion} precoding conditioned aggregation beamforming is shown to be optimal as follows. 
\begin{lemma}[Optimal Sensor Precoder]\label{MISO_ZF} \emph{
For the MISO case, given a data aggregation beamformer $\ba$, the computation error is minimized by the following channel-inversion precoders at sensors:
\begin{equation}\label{Eq:MISO_ZF}
b_k^* = \frac{1}{\sqrt{\eta} \bold{f}^H\bold{h}_k}, \qquad \forall k,
\end{equation}
where $\bold{f}$ is defined by normalizing the aggregation beamformer, and $\eta \in \mathbb{R}_{+}$ is chosen to satisfy the sensor transmission power constraints in Problem P2, i.e., to guarantee that $\bold{a} = \sqrt{\eta} \bold{f}$.
}
\end{lemma}
\proof
See Appendix \ref{App:MISO_ZF}.
\endproof

By substituting the result in Lemma \ref{MISO_ZF}, Problem P2 can be transformed into the equivalent problem of minimizing the variable $\eta$, called the denoising factor:
\begin{equation*}\textbf{(P3)}\qquad
\begin{aligned}
\min_{\eta, \bold{f}, \{\bold{u}_k\}, \{P_k\}} \quad 
&\eta\\ 
\text{s.t.} \quad 
&\frac{1}{\eta\|\bold{h}_k^H\bold{f}\|^2} \le \gamma_k \|\bold{u}_k^H\bold{g}_k\|^2 P_k,~\forall k,\\
&\sum_{k=1}^{K} P_k \le P_0,~\bold{f}^H \bold{f} = 1,\\
&\bold{u}_k^H \bold{u}_k = 1,~\forall k.
\end{aligned}
\end{equation*}
The name of $\eta$ is given based on the fact that reducing it suppresses the effect of channel noise by increasing symbol energy (or equivalently $\{b_n^*\}$ as observed from \eqref{Eq:MISO_ZF}). Since minimizing $\eta$ is equivalent to maximizing $\frac{1}{\eta}$, one can incorporate the power constraint into the objective function of Problem P3 and have the following equivalent \emph{max-min} problem:

\begin{equation*}\textbf{(P4)}\qquad 
\begin{aligned}
\max_{\bold{f}, \{\bold{u}_k\}, \{P_k\}} \quad 
&\min_{k} \gamma_k \|\bold{u}_k^H\bold{g}_k\|^2 \|\bold{h}_k^H\bold{f}\|^2 P_k\\ 
\text{s.t.} \quad 
&\sum_{k=1}^{K} P_k \le P_0,\\
&\bold{f}^H \bold{f} = 1,\\
&\bold{u}_k^H \bold{u}_k = 1,~\forall k.
\end{aligned}
\end{equation*}
Thought Problem P4 has a simpler form than Problem P3, it remains non-convex. To tackle the challenge, we recast the problem into an equivalent \emph{outer-inner sub-problems formulation}, having the following form:
\begin{equation*}
\boxed{
\textbf{(P5)}\qquad \underbrace{\max_{\bold{f}^H\bold{f}=1} \quad \underbrace{ \left\{
\begin{aligned}
\max_{\{\bold{u}_k\}, \{P_k\}} \quad 
&\min_{k} \gamma_k \|\bold{u}_k^H\bold{g}_k\|^2 \|\bold{h}_k^H\bold{f}\|^2 P_k\\ 
\text{s.t.} \quad 
&\sum_{k=1}^{K} P_k \le P_0,\\
&\bold{u}_k^H\bold{u}_k=1,~\forall k.
\end{aligned}
\right\}}_{\textbf{inner sub-problem}}
}_{\textbf{outer sub-problem}}.
}
\end{equation*}

The inner sub-problem solves the optimal power allocation and energy beamforming, while the outer sub-problem aims at further optimizing the aggregation beamformer. The two sub-problems are tractable and solved separately in the following sub-sections.

\subsection{Optimal Wireless Power Control and Energy Beamforming}
A close observation of the inner sub-problem in P5 indicates that the optimal energy beamforming can be firstly solved in closed form, based on which the optimal power allocation can be then derived as a function of the normalized  aggregation beamformer $\bold{f}$. The first key result of this sub-section is presented as follows.

\begin{proposition}[Optimal Energy Transmit Beamformer]\label{MISO_Beam} \emph{Consider the MISO case. For each sensor, the optimal power beam should point to the direction of the corresponding WPT channel, i.e., 
\begin{equation}\label{Eq:MISO_Beam}
\bold{u}_k^* = \frac{\bold{g}_k}{\|\bold{g}_k\|}.
\end{equation}
}
\end{proposition}
\proof
In the inner sub-problem of P5, $\|\bold{u}_k^H\bold{g}_k\|^2 \le \|\bold{u}_k\|^2 \|\bold{g}_k\|^2$ with the equality holds if and only if $\bold{u}_k$ has the same direction as $\bold{g}_k$.
\endproof

By substituting \eqref{Eq:MISO_Beam}, the inner sub-problem is transformed into the following problem:
\begin{equation*}\textbf{(P6)}\qquad 
\begin{aligned}
\max_{\{P_k\}} \quad 
&\min_{k} \gamma_k \|\bold{g}_k\|^2 \|\bold{h}_k^H\bold{f}\|^2 P_k\\ 
\text{s.t.} \quad 
&\sum_{k=1}^{K} P_k \le P_0.
\end{aligned}
\end{equation*}
To solve Problem P6, a necessary condition for the optimal solution is derived as follows.
\begin{lemma}[Optimal Wireless Power Control]\label{MISO_Equal} \emph{For the MISO case, given the aggregation beamformer $\bold{f}$, the optimal power control should have the following form:
\begin{equation}\label{Eq:MISO_Const}
P_k^* = \frac{C}{\gamma_k \|\bold{h}_k^H\bold{f}\|^2 \|\bold{g}_k\|^2},~\forall k,
\end{equation}
where $C$ is some constant.
}
\end{lemma}
\proof
See Appendix \ref{App:MISO_Equal}.
\endproof

Based on Lemma \ref{MISO_Equal}, Problem P6 reduces to the following:
\begin{equation*}\textbf{(P7)}\qquad 
\begin{aligned}
\max_{C} \quad 
&C\\ 
\text{s.t.} \qquad 
&\sum_{k=1}^{K} \frac{C}{\gamma_k \|\bold{h}_k^H\bold{f}\|^2 \|\bold{g}_k\|^2} \le P_0.
\end{aligned}
\end{equation*}
Problem P7 is trivial and the solution is
\begin{equation}\label{Eq:MISO_C}
C^*=\frac{P_0}{\sum_{k=1}^{K} \frac{1}{\gamma_k \|\bold{h}_k^H\bold{f}\|^2 \|\bold{g}_k\|^2}}.
\end{equation}
Then combining \eqref{Eq:MISO_Const} and \eqref{Eq:MISO_C} gives the following second key result of this sub-section.
\begin{proposition}[Optimal Power Allocation]\label{MISO_Power} \emph{For the MISO case, given the aggregation beamformer $\bold{f}$, the optimal power allocation is given by
\begin{equation}\label{Eq:MISO_Power}
P_k^* = \frac{P_0}{{\gamma_k\|\bold{h}_k^H\bold{f}\|^2\|\bold{g}_k\|^2 \sum_{k=1}^{K} \frac{1}{\gamma_k \|\bold{h}_k^H\bold{f}\|^2 \|\bold{g}_k\|^2}}}.
\end{equation}
}
\end{proposition}

\begin{remark}[Optimal WPT Strategies]\emph{
It can be observed form Proposition \ref{MISO_Beam} that the optimal power beam points in the direction of its corresponding WPT channel. With the beams thus steered, the optimal power allocation attempts to equalize the effective close-loop channels of different sensors that cascade the WPT channels specified by the gains $\|\bold{g}_k\|^2$ and AirComp channels specified by the gains $\|\bold{h}_k^H \bold{f} \|^2$.}
\end{remark}

\begin{remark}[Sensor Scheduling]\emph{
The current design focuses on a set of sensors selected for WDA. The issue of scheduling is not addressed but important. To be specific, deep fading in a cascading WPT-AirComp channel will lead to highly noisy data collected from the corresponding sensor and thereby amplify the computation error. This is reflected in high transmit power for inverting a poor channel (see \eqref{Eq:MISO_Power}) and thereby reduces the average SNR of aggregated data. Thus, it is important to select sensors with both reliable WPT and AirComp channels.}
\end{remark}

\subsection{Optimal Aggregation Beamforming}
Given the solution of the inner sub-problem in the preceding sub-section, the outer sub-problem in P5 for aggregation beamformer optimization can be simplified by substituting the optimal energy beamformer in \eqref{Eq:MISO_Beam} and optimal power allocation in \eqref{Eq:MISO_Power}:
\begin{equation*}\textbf{(P8)}\qquad 
\begin{aligned}
\max_{\bold{f}} \quad 
&\frac{P_0}{\sum_{k=1}^{K}\frac{1}{\gamma_k \|\bold{f}^H\bold{h}_k\|^2 \|\bold{g}_k\|^2}}\\ 
\text{s.t.} \quad 
&\bold{f}^H\bold{f}=1.
\end{aligned}
\end{equation*}

Note that $\|\bold{f}^H\bold{h}_k\|^2=\tr(\bold{h}_k\bold{h}_k^H\bold{f}\bold{f}^H)$, Problem P8 is equivalent to the following problem:
\begin{equation*}\textbf{(P9)}\qquad
\begin{aligned}
\min_{\bold{f}} \quad 
&\sum_{k=1}^{K} \frac{1}{\gamma_k \tr(\bold{h}_k\bold{h}_k^H\bold{f}\bold{f}^H)\|\bold{g}_k\|^2 P_0}\\ 
\text{s.t.} \quad
&\bold{f}^H\bold{f}=1.
\end{aligned}
\end{equation*}
Though having a simple structure, Problem P9 is still challenging due to the non-convex norm constraint on $\bold{f}$. To tackle the constraint, the celebrated SDR technique is applied where the non-convex constraint in Problem P9 is relaxed by its convex hull. 
\begin{lemma}[Convex Hull Relaxation \cite{overton1992sum}]\label{Hull} \emph{Let $\Omega_1 = \{\bold{X}:\bold{X}=\bold{W}\bold{W}^H, \bold{W}^H\bold{W}=\bold{I}_d\}$ and $\Omega_2 = \{\bold{X}:\tr(\bold{X})=d, 0 \preceq \bold{X} \preceq \bold{I}\}$, wherein $\bold{W}$ is of the size $m$ by $d$ and $\bold{X}$ has the dimension of $m$ by $m$. The second condition $0 \preceq \bold{X} \preceq \bold{I}$ means that both $\bold{X}$ and $\bold{I}-\bold{X}$ are positive semi-definite. Then, $\Omega_2$ is the convex hull of $\Omega_1$, and $\Omega_1$ is the set of extreme points of $\Omega_2$.}
\end{lemma}

Thereby relaxing Problem P9 gives:
\begin{equation*} \textbf{(P10)}\qquad
\begin{aligned} 
\min_{\hat{\bold{F}}} \quad 
&\sum_{k=1}^{K} \frac{1}{\gamma_k \tr(\bold{h}_k\bold{h}_k^H\hat{\bold{F}})\|\bold{g}_k\|^2 P_0}\\ 
 \text{s.t.} \quad
&\tr(\hat{\bold{F}})=1,~ 0 \preceq \hat{\bold{F}}\preceq\bold{I},
\end{aligned}
\end{equation*}
where $\hat{\bold{F}}=\bold{f}\bold{f}^H$. Then, the convexity of Problem P10 is established in the following lemma.

\begin{lemma}[Convexity of Problem P10]\label{MISO_Conv} \emph{
Problem P10 is a convex problem.}
\end{lemma}
\proof
See Appendix \ref{App:MISO_Conv}.
\endproof

Upon solving the Problem P10 via a convex problem solver (e.g., the cvx toolbox in MATLAB) and attaining the globally optimal solution $\hat{\bold{F}}^*$, the next task is to retrieve from it a feasible solution to Problem P9 denoted by $\tilde{\bold{f}}$. Since the rank of $\hat{\bold{F}}^*$ might be larger than one, the Gaussian randomization algorithm proposed in \cite{luo2010semidefinite} can be applied to extract $\tilde{\bold{f}}$ from $\hat{\bold{F}}^*$. The main procedure of applying the algorithm to the current design is summarized in Algorithm~\ref{Al:MISO}.

\begin{remark}[Optimality of SDR Solution]\emph{
It is worth pointing out that the SDR technique can lead to an optimal solution. If a rank one $\hat{\bold{F}}^*$ is solved, the global optimal solution can be immediately achieved by extracting the dominant eigenvector of $\hat{\bold{F}}^*$ as the normalized data aggregation beamformer, i.e., $\bold{f}^* = [\bold{V}_{\hat{\bold{F}}}]_{:,1}$. As shown in the simulation later, $\hat{\bold{F}}^*$ has a high probability to be rank one. 
}
\end{remark}

\begin{algorithm}[tt]
\caption{Gaussian Randomization Algorithm for MISO WP-AirComp}
\label{Al:MISO}
\begin{itemize}
\item{\textbf{Initialization}: Given an SDR solution $\hat{\bold{F}}^*$, and the number of random samples $M$.}
\item{\textbf{Gaussian Random Sampling}: \\
(1) Perform eigen decomposition $[\bold{V}_{\hat{\bold{F}}}, \bold{\Sigma}_{\hat{\bold{F}}}] = \text{eig}(\hat{\bold{F}}^*)$.\\
(2) Generate $M$ random vectors $\bold{z}_{m} \sim \mathcal{CN}(\bold{0},\bold{I})$ with $\bold{0} \in \mathbb{C}^{N_{\sf AP} \times 1}$ and $\bold{I} \in \mathbb{C}^{N_{\sf AP} \times N_{\sf AP}}$.\\
(3) Retrieve $M$ feasible solutions $\{\bold{f}_{m}\}$ from $\{\bold{z}_{m}\}$ via  $\bold{f}_{m} = \frac{\bold{V}_{\hat{\bold{F}}} \bold{\Sigma}_{\hat{\bold{F}}}^{1/2} \bold{z}_m^H}{\|\bold{V}_{\hat{\bold{F}}} \bold{\Sigma}_{\hat{\bold{F}}}^{1/2} \bold{z}_m^H\|}$, $m = 1,...,M$\\
(4) Select the best $\bold{f}_{m}$ that leads to the minimum objective, namely $\bold{f}_{m}^* = \arg \min_{m}\sum_{k=1}^{K} \frac{1}{\gamma_k \tr(\bold{h}_k\bold{h}_k^H\bold{f}_{m}\bold{f}_{m}^H)\|\bold{g}_k\|^2 P_0}$.\\
(5) Output $\tilde{\bold{f}}=\bold{f}_{m}^*$ as the approximated optimal normalized aggregation beamformer.}
\end{itemize}
\end{algorithm}

\section{Joint Design for WPT and AirComp: MIMO Case}
In the preceding section, we consider the MISO case with single-antenna sensors. In this section, the results are extended to the general MIMO case with multi-antenna sensors. In particular, the original vector-function WP-AirComp problem in Problem P1 is solved. The solution builds on the outer-inner framework developed in the previous section. 

To further develop the framework for the MIMO case, the non-convex Problem P1 is first simplified by showing the optimality of the zero-forcing (channel inversion) precoder conditioned on the aggregation beamformer as follows. 
\begin{lemma}[Optimal MIMO Precoder]\label{MIMO_ZF} \emph{
Given an aggregation beamformer $\bold{A}$, the computation error is minimized by the following zero-forcing precoders:
\begin{equation}\label{Eq:MIMO_ZF}
\bold{B}_k^* = (\bold{A}^H \bold{H}_k)^H (\bold{A}^H \bold{H}_k \bold{H}_k^H \bold{A})^{-1}, \qquad \forall k.
\end{equation}
}
\end{lemma}

The proof of Lemma \ref{MIMO_ZF} is similar to that of Lemma \ref{MISO_ZF} shown in Appendix \ref{App:MISO_ZF}, thus omitted for brevity. Let $\bold{F}$ denote the normalized aggregation beamformer with $\tr(\bold{F}\bold{F}^H) = 1$ and thus $\bold{A} = \sqrt{\eta} \bold{F}$ with $\eta$ being the norm of $\bold{A}$.
Then Problem P1 can be reduced to the following form by substituting \eqref{Eq:MIMO_ZF}:
\begin{equation*}\textbf{(P11)}\qquad 
\begin{aligned}
\min_{\eta, \bold{F}, \{\bold{u}_k\}, \{\bold{v}_k\},\{P_k\}} \quad 
&\eta\\ 
\text{s.t.} \quad 
&\frac{1}{\eta}\tr\l((\bold{F}^H\bold{H}_k\bold{H}_k^H\bold{F})^{-1}\r) \le \gamma_k \|\bold{u}_k^H\bold{G}_k\bold{v}_k\|^2 P_k,~\forall k,\\
&\sum_{k=1}^{K} P_k \le P_0,\\
&\tr(\bold{F}\bold{F}^H) = 1,\\
&\bold{u}_k^H\bold{u}_k=1,~\forall k,\\
&\bold{v}_k^H\bold{v}_k=1,~\forall k.
\end{aligned}
\end{equation*}
Note that the first set of power constraints in Problem P11 can be rewritten as:
\begin{equation}\label{Eq:MIMO_Constraint}
\frac{1}{\eta} \le \min_{k} \frac{\gamma_k \|\bold{u}_k^H\bold{G}_k\bold{v}_k\|^2 P_k}{\tr\l((\bold{F}^H\bold{H}_k\bold{H}_k^H\bold{F})^{-1}\r)}.
\end{equation}
Note that the minimum $\eta$ in Problem P2 is achieved when the above constraint is active (i.e., the equality holds). Since minimizing $\eta$ is equivalent to maximizing $\frac{1}{\eta}$, one can move the power constraint to the objective function and have the following equivalent \emph{max-min} problem:
\begin{equation*}\textbf{(P12)}\qquad
\begin{aligned}
\max_{\bold{F}, \{\bold{u}_k\}, \{\bold{v}_k\},\{P_k\}} \quad 
&\min_k \frac{\gamma_k \|\bold{u}_k^H\bold{G}_k\bold{v}_k\|^2 P_k}{\tr\l((\bold{F}^H\bold{H}_k\bold{H}_k^H\bold{F})^{-1}\r)}\\ 
\text{s.t.} \quad 
&\sum_{k=1}^{K} P_k \le P_0,\\
&\tr(\bold{F}\bold{F}^H) = 1,\\
&\bold{u}_k^H\bold{u}_k=1,~\forall k,\\
&\bold{v}_k^H\bold{v}_k=1,~\forall k.
\end{aligned}
\end{equation*}

Similar to the MISO counterpart, Problem P12 can be recast as the following outer-inner sub-problems: 
\begin{equation*}
\boxed{
\textbf{(P13)}\qquad \underbrace{\max_{\tr(\bold{F}\bold{F}^H) = 1} \quad \underbrace{ \left\{
\begin{aligned}
\max_{\{\bold{u}_k\}, \{\bold{v}_k\},\{P_k\}} \quad 
&\min_{k} \frac{\gamma_k \|\bold{u}_k^H\bold{G}_k\bold{v}_k\|^2 P_k}{\tr\l((\bold{F}^H\bold{H}_k\bold{H}_k^H\bold{F})^{-1}\r)}\\ 
\text{s.t.} \quad 
&\sum_{k=1}^{K} P_k \le P_0,\\
&\bold{u}_k^H\bold{u}_k=1,~\forall k,\\
&\bold{v}_k^H\bold{v}_k=1,~\forall k,
\end{aligned}
\right\}}_{\textbf{inner sub-problem}}
}_{\textbf{outer sub-problem}}
}
\end{equation*}
where the inner sub-problem contains MIMO power control and energy beamforming optimization, and the MIMO aggregation beamformer design gives the outer sub-problem.

\subsection{Optimal Wireless Power Control and Energy Beamforming}
We firstly tackle the inner sub-problem in Problem P13. The optimal energy transmit and receive beamformers are solved as a function of the normalized aggregation beamformer $\bold{F}$ as shown below.
\begin{proposition}[Optimal MIMO Energy Beamformers]\label{MIMO_Beam} \emph{
For each sensor, the optimal energy transmit/receive beamformers point to the left/right dominant singular vector of the WPT channel matrix, namely, 
\begin{equation}\label{Eq:MIMO_Beam}
\bold{u}_k^* = \frac{\bold{u}_{\max}(\bold{G}_k)}{\|\bold{u}_{\max}(\bold{G}_k)\|}~\text{and}~\bold{v}_k^* = \frac{\bold{v}_{\max}(\bold{G}_k)}{\|\bold{v}_{\max}(\bold{G}_k)\|},
\end{equation}
where $\bold{u}_{\max}(\bold{G}_k)$ and $\bold{v}_{\max}(\bold{G}_k)$ denote the left and right dominant singular vectors of matrix $\bold{G}_k$, respectively.
}
\end{proposition}
\proof
Starting from the well-known \emph{Rayleigh-quotient} inequality
\begin{align}
\|\bold{u}_{k}^H\bold{G}_{k}\bold{v}_{k}\|^2 \le \|\bold{u}_{k}\|^2 \sigma_{\max}^2(\bold{G}_k) \|\bold{v}_{k}\|^2,
\end{align}
where the equality holds if and only if $\bold{u}_k$ and $\bold{v}_k$ have the same direction with the left and right dominant singular vector $\bold{u}_{\max}(\bold{G}_k)$ and $\bold{v}_{\max}(\bold{G}_k)$. Therefore, one can readily note that the objective in Problem P13 can be maximized with respect to $\bold{u}_k$ and $\bold{v}_k$ by setting them as shown in \eqref{Eq:MIMO_Beam}, which completes the proof.
\endproof

By substituting \eqref{Eq:MIMO_Beam} in Lemma~\ref{MIMO_Beam}, the inner sub-problem is transformed into the following problem:
\begin{equation*}\textbf{(P14)}\qquad 
\begin{aligned}
\max_{\{P_k\}} \quad 
&\min_{k} \frac{\gamma_k \sigma_{\max}^2(\bold{G}_k) P_k}{\tr\l((\bold{F}^H\bold{H}_k\bold{H}_k^H\bold{F})^{-1}\r)}\\ 
\text{s.t.} \quad 
&\sum_{k=1}^{K} P_k \le P_0, 
\end{aligned}
\end{equation*}
where $\sigma_{\max}(\bold{G}_k)$ denotes the dominant singular value of the WPT channel matrix $\bold{G}_k$. To solve Problem P14, a necessary condition for the optimal solution is provided as follows.
\begin{lemma}[Optimal Wireless Power Control]\label{MIMO_Equal} \emph{For the MIMO case, the optimal power control should have the following form:
\begin{equation}\label{Eq:MIMO_Const}
P_k^* = \frac{C\tr\l((\bold{F}^H\bold{H}_k\bold{H}_k^H\bold{F})^{-1}\r)}{\gamma_k \sigma_{\max}^2(\bold{G}_k)},~\forall k,
\end{equation}
where $C$ is some constant.
}
\end{lemma}
\proof
See Appendix \ref{App:MIMO_Equal}.
\endproof

Based on Lemma \ref{MIMO_Equal}, Problem P14 reduces to the following:
\begin{equation*}\textbf{(P15)}\qquad 
\begin{aligned}
\max_{C} \quad 
&C\\ 
\text{s.t.} \quad 
&\sum_{k=1}^{K} \frac{C\tr\l((\bold{F}^H\bold{H}_k\bold{H}_k^H\bold{F})^{-1}\r)}{\gamma_k \sigma_{\max}^2(\bold{G}_k)} \le P_0.
\end{aligned}
\end{equation*}
Problem P15 can be easily solved with the optimal solution:
\begin{align}
C^*=\frac{P_0}{\sum_{k=1}^{K} \frac{\tr\l((\bold{F}^H\bold{H}_k\bold{H}_k^H\bold{F})^{-1}\r)}{\gamma_k \sigma_{\max}^2(\bold{G}_k)}}.
\end{align} 
Thus the optimal power control policy is given as follows.

\begin{proposition}[Optimal MIMO Power Allocation]\label{MIMO_Power} \emph{For the MIMO case, the optimal power allocation is given by
\begin{equation}\label{Eq:MIMO_Power}
P_k^* = \frac{P_0\tr\l((\bold{F}^H\bold{H}_k\bold{H}_k^H\bold{F})^{-1}\r)}{\gamma_k \sigma_{\max}^2(\bold{G}_k) \sum_{k=1}^{K} \frac{\tr\l((\bold{F}^H\bold{H}_k\bold{H}_k^H\bold{F})^{-1}\r)}{\gamma_k \sigma_{\max}^2(\bold{G}_k)}}.
\end{equation}
}
\end{proposition}

\begin{remark}[Optimal WPT for MIMO Case]\emph{
Similar to its MISO counterpart, \eqref{Eq:MIMO_Power} indicates that the optimal transmit beam points to the left dominant eigen-direction of the WPT channel matrix, and the optimal receive beam to the right dominant eigen-direction. Moreover, the allocated power to a sensor $k$ is inversely proportional to its effective close-loop channel gain that combines the dominant singular value of WPT channel matrix $\bold{G}_k$ and  the beamed AirComp channel after aggregation beamforming, i.e., $\frac{1}{\tr\l({ (\bold{F}^H\bold{H}_k\bold{H}_k^H\bold{F})^{-1}}\r) }$ (see Proposition \ref{MIMO_Power}).
}
\end{remark}

\subsection{Optimal Aggregation Beamformer}
Given the solution of the inner sub-problem presented above, the outer sub-problem in P13 for aggregation beamformer optimization can be obtained by substituting the optimal WPT strategies in \eqref{Eq:MIMO_Beam} and  \eqref{Eq:MIMO_Power} into Problem P13. It follows that:
\begin{equation*}\textbf{(P16)}\qquad 
\begin{aligned}
\min_{\bold{F}} \quad 
&\sum_{k=1}^{K} \frac{\tr\l((\bold{F}^H\bold{H}_k\bold{H}_k^H\bold{F})^{-1}\r)}{\gamma_k \sigma_{\max}^2(\bold{G}_k) P_0}\\ 
\text{s.t.} \quad 
&\tr(\bold{F}\bold{F}^H) = 1.
\end{aligned}
\end{equation*}
Problem P16 is difficult to solve due to the non-convex objective involving $\tr((\bold{F}^H\bold{H}_k\bold{H}_k^H\bold{F})^{-1})$. To overcome the difficulty, we adopt the following inequality relaxation of the objective function:
\begin{equation}\label{ineq}
\tr((\bold{F}^H\bold{H}_k\bold{H}_k^H\bold{F})^{-1}) \le \frac{L}{\lambda_{\min}(\bold{H}_k^H\bold{F}\bold{F}^H\bold{H}_k)},
\end{equation}
where the equality holds given a well-conditioned channel, i.e., the singular values of $\bold{H}_k$ are identical. Using \eqref{ineq}, a relaxed version of Problem P16 is posed as follows:
\begin{equation*}\textbf{(P17)}\qquad 
\begin{aligned}
\min_{\bold{F}} \quad 
&\sum_{k=1}^{K}\frac{L}{\gamma_k \sigma_{\max}^2(\bold{G}_k)\lambda_{\min}(\bold{H}_k^H\bold{F}\bold{F}^H\bold{H}_k)P_0}\\ 
\text{s.t.} \quad 
&\tr(\bold{F}\bold{F}^H) = 1.
\end{aligned}
\end{equation*}

According to \cite{luo2010semidefinite}, by introducing a new variable $\hat{\bold{F}} = \bold{F}\bold{F}^H$, an equivalent formulation of Problem P17 is obtained as follows:
\begin{equation*}\textbf{(P18)}\qquad
\begin{aligned}
\min_{\bold{F}} \quad 
&\sum_{k=1}^{K}\frac{L}{\gamma_k \sigma_{\max}^2(\bold{G}_k)\lambda_{\min}(\bold{H}_k^H\hat{\bold{F}}\bold{H}_k)P_0}\\ 
\text{s.t.} \quad 
&\tr(\hat{\bold{F}})=1,~\rank(\hat{\bold{F}}) = L,~\hat{\bold{F}} \succeq 0.
\end{aligned}
\end{equation*}
It can be observed that the only non-convex constraint in Problem P18 is $\rank(\hat{\bold{F}}) = L$, thus we may remove it to obtain the following relaxed version:
\begin{equation*}\textbf{(P19)}\qquad
\begin{aligned}
\min_{\bold{F}} \quad 
&\sum_{k=1}^{K}\frac{L}{\gamma_k \sigma_{\max}^2(\bold{G}_k)\lambda_{\min}(\bold{H}_k^H\hat{\bold{F}}\bold{H}_k)P_0}\\ 
\text{s.t.} \quad 
&\tr(\hat{\bold{F}})=1,~\hat{\bold{F}} \succeq 0.
\end{aligned}
\end{equation*}

\noindent The convexity of Problem P19 is shown in the following lemma.
\begin{lemma}[Convexity of Problem P19]\label{MIMO_Conv} \emph{
Problem P19 is a convex problem.}
\end{lemma}
\proof
See Appendix \ref{App:MIMO_Conv}.
\endproof

\begin{algorithm}[tt]
\caption{Gaussian Randomization Algorithm for MIMO WP-AirComp}
\label{Al:MIMO}
\begin{itemize}
\item{{\bf Initialization}: Given an SDR solution $\hat{\bold{F}}^*$, and the number of random samples $M$.}
\item{{\bf Gaussian Random Sampling}: \\
(1) Perform eigen decomposition $[\bold{V}_{\hat{\bold{F}}}, \bold{\Sigma}_{\hat{\bold{F}}}] = \text{eig}(\hat{\bold{F}}^*)$.\\
(2) Generate $M$ random matrices $\bold{Z}_m \sim \mathcal{CN}(\bold{0},\bold{I})$ with $\bold{Z}_m \in \mathbb{C}^{N_{\sf AP} \times L}$, $\bold{0} \in \mathbb{C}^{N_{\sf AP} \times L}$ and $\bold{I} \in \mathbb{C}^{N_{\sf AP}\times N_{\sf AP}}$.\\
(3) Retrieve $M$ feasible solutions $\{\bold{F}_{m}\}$ from $\{\bold{Z}_m\}$  by extracting the $L$ left dominant singular vectors from $\bold{V}_{\hat{\bold{F}}} \bold{\Sigma}_{\hat{\bold{F}}}^{1/2} \bold{Z}_m^H$, denoted by $\bold{V}_m$, and normalizing it by its norm $\sqrt{L}$, namely $\bold{F}_{m} = \frac{1}{\sqrt{L}} \bold{V}_m$, such that the constraint $\tr (\bold{F}_{m} \bold{F}_{m}^H) = 1$ can be enforced.\\
(4) Select the best $\bold{F}_{m}$ that leads to the minimum objective, namely $\bold{F}_{m}^* = \arg \min_{m}\sum_{k=1}^{K}\frac{LT}{\gamma_k \sigma_{\max}^2(\bold{G}_k)\lambda_{\min}(\bold{H}_k^H\bold{F}_{m}\bold{F}_{m}^{H}\bold{H}_k)P_0 T_0}$.\\
(5) Output $\tilde{\bold{F}}=\bold{F}_{m}^*$ as the approximated optimal normalized aggregation beamformer.}
\end{itemize}
\end{algorithm}

Upon attaining the globally optimal solution of Problem P19, denoted by $\hat{\bold{F}}^*$, the remaining task is to convert it into a feasible solution of Problem P16, denoted by $\tilde{\bold{F}}$, of rank $L$. To this end, the Gaussian randomization algorithm for the MISO case in Algorithm~\ref{Al:MISO} is generalized to the MIMO case for searching the close-to-optimal approximate solution for Problem P16, as summarized in Algorithm~\ref{Al:MIMO}. 

\section{Further Discussion}
In this section, we provide further discussion to gain more insights into the properties of the WP-AirComp design in the preceding sections.

\subsection{Computation Error Minimization versus SNR Maximization}
An intuitive and alternative design criterion for WP-AirComp could be one that maximizes the (total) receive SNR at the server. This criterion, however, leads to a completely different strategy from the counterpart that minimizes the computation error, as shown in the sequel. To clarify this point, we consider the simple MISO case for example. Let $\rho$ denote the receive SNR and it can be defined based on the channel model in \eqref{estimated_s} as follows: 
\begin{align*}
\rho &= \sum_{k=1}^{K} \frac{\|b_k\|^2 \bold{a}^H\bold{h}_k \bold{h}_k^H\bold{a}}{\sigma_n^2\bold{a}^H\bold{a}}\\
& = \frac{1}{\sigma_n^2} \sum_{k=1}^K  \gamma_k \|\bold{u}_k^H\bold{g}_k\|^2 \|\bold{h}_k^H\bold{f}\|^2 P_k,
\end{align*}
where the second equality is attained by substituting $\bold{a} = \sqrt{\eta} \bold{f}$, and the power constraint $\|b_k\|^2 =\gamma_k \|\bold{u}_k^H\bold{g}_k\|^2 P_k$. Thus, the SNR maximization problem can be casted as:

\begin{equation*}(\textbf{Max SNR})\qquad 
\begin{aligned}
\max_{\bold{f}, \{\bold{u}_k\}, \{P_k\}} \quad 
&\sum_{k=1}^K \gamma_k \|\bold{u}_k^H\bold{g}_k\|^2 \|\bold{h}_k^H\bold{f}\|^2 P_k\\ 
\text{s.t.} \quad 
&\sum_{k=1}^{K} P_k \le P_0,\\
&\bold{f}^H \bold{f} = 1,\\
&\bold{u}_k^H \bold{u}_k = 1,~\forall k.
\end{aligned}
\end{equation*}
Although the above problem and P4 differ slightly only in the objective function, the resultant resource allocation strategies are fundamentally different. The strategy from SNR maximization tends to allocate more power to a sensor with better channel condition so as to maximize the sum effective channel gain. In contrast, the strategy from computation error minimization in Problem P4 attempts to equalize effective channel gain across different sensors (see Proposition~\ref{MISO_Power}). This suggests the fundamental difference between WDA and conventional data communication.

\subsection{How does WPT Help AirComp?}
As mentioned, wireless power control provides an additional design dimension for reducing the AirComp error. In the sequel, we provide insights to understand the performance gain from relevant design presented in the preceding sections. To help exposition, a benchmark scheme is considered that equally allocates wireless power, i.e., $P_k = \frac{P_0}{K}$. Consider the aggregation beamformer design problem in P4. Note that under equal power allocation, P4 is converted to a NP-hard problem as shown below:
\begin{equation*}{(\tilde{\textbf{P}})} \qquad
\begin{aligned}
\min_{\bold{f}} \quad 
&\max_{k} \frac{K}{\gamma_k \|\bold{g}_k\|^2 \|\bold{h}_k^H\bold{f}\|^2 P_0}\\ 
\text{s.t.} \quad 
&\bold{f}^H \bold{f} = 1.
\end{aligned}
\end{equation*}
Comparing the objective functions of Problems $\tilde{\textbf{P}}$ and P9 (with wireless power control) gives:
\begin{align}\label{max_to_sum}
\sum_{k=1}^{K} \frac{1}{\gamma_k \|\bold{g}_k\|^2 \|\bold{h}_k^H\bold{f}\|^2 P_0} \le \max_{k} \frac{K}{\gamma_k \|\bold{g}_k\|^2 \|\bold{h}_k^H\bold{f}\|^2 P_0}.
\end{align}
In other words, wireless power control reduces the computation error.

\section{Simulation}
In this section, the performance of our proposed WP-AirComp framework is evaluated by simulation. The performance metric is the normalized computation error, defined by $\text{MSE}/K$ with MSE defined in \eqref{MSE_function}.  The simulation parameters are set as follows unless specified otherwise. The number of sensors is $K=5$.  In the MISO case, the number of antennas at the server is set as $N_{\sf AP} = 20$. The number of computed functions is set to be $L=1$ as only $1$ antenna is available at sensor side. In the MIMO case, the antenna setting at the server is given by $N_{\sf AP} = 30$ , while at sensor side we assume $N_{\sf SN} = 5$ antennas for energy receive beamforming or data precoding, which is equal to the number of computed function $L=5$. All the WPT and Aircomp channels are assumed to be i.i.d. \emph{Rician fading}, modeled as i.i.d. complex Gaussian random variables with non-zero mean $\mu = 1$ and variance $\sigma^2 = 1$. In addition, the maximum transmission power is set as $P_0 = 1$. The effective power conversion efficiency follows a uniform distribution with $\eta_n \in (0,1)$, and the noise variance is assumed to be $1$.

\subsection{Baseline Schemes}
We consider three baseline schemes designed based on two classic approaches: \emph{antenna selection} (AS) and \emph{eigenmode beamforming} (EB). All three schemes assume the channel-inversion data precoding in \eqref{Eq:MISO_ZF} or \eqref{Eq:MIMO_ZF} and also the optimal energy beamforming in \eqref{Eq:MISO_Beam} or \eqref{Eq:MIMO_Beam} depending on whether MISO or MIMO case is considered. The difference between the three schemes lie in the aggregation beamformer and the wireless power allocation policy. Define the sum-channel matrix $\bold{H}_{\sf sum} = \sum_{k=1}^K \bold{H}_k$. For the first two baseline schemes, the optimal power allocation is used by solving the inner sub-problems of Problem P5 and P13 under the condition that the aggregation beamformer is set to be AS or EB to enhance the receive SNR. The AS scheme selects the $L$ receive antenna observing the largest channel gains in the sum channel $\bold{H}_{\sf sum}$, while the EB scheme selects the $L$ largest eigenmode of $\bold{H}_{\sf sum}$ for AirComp  and thus consists of the $L$ dominant left eigenvectors of $\bold{H}_{\sf sum}$. The third baseline scheme assumes equal power allocation $P_k = \frac{P_0}{K}$ and solves the resultant outer sub-problems of Problem P5 and P13 to attain the aggregation beamformer. For fair comparison, all aggregation beamformers in the baseline schemes are scaled to have the same norm.

\begin{table}[t]
\centering
\caption{Probability of SDR solution being optimal.}
\begin{tabular}{|c|c|c|c|c|c|c|c|c|c|c|}
\hline
$N_{\sf AP}$ & $5$ & $ 10$ & $15$ & $ 20$ & $25$ & $30$ & $35$ & $40$ & $45$ & $50$ \\ 
\hline
$P_{\sf opt}$ & $85.77\%$ & $86.03\%$ & $86.10\%$ & $86.71\%$ & $86.96\%$ & $87.31\%$ & $87.69\%$ & $88.15\%$ & $88.16\%$ & $88.18\%$ \\ 
\hline
\end{tabular}
\label{summary:table2}
\end{table}

\subsection{Optimality of the SDR Solution}
Consider the MISO case in Section IV, the probability of SDR solution for Problem P10 to be globally optimal, i.e., $\rank(\hat{\bold{F}}^*) = 1$, is evaluated under various antenna settings at the server as summarized in Table \ref{summary:table2}. It can be observed that the probability of SDR solution to be optimal is more than $85\%$ for different settings. Moreover, the probability is observed to increase with the growth of the antenna numbers at the server, which implies that equipping the server with more antennas will boost the AirComp accuracy to some extent.

\subsection{Computation Performance of WP-AirComp}
\begin{figure}[t]
  \centering
  \subfigure[MISO case]{
  \label{MISO_P}
  \includegraphics[scale=0.38]{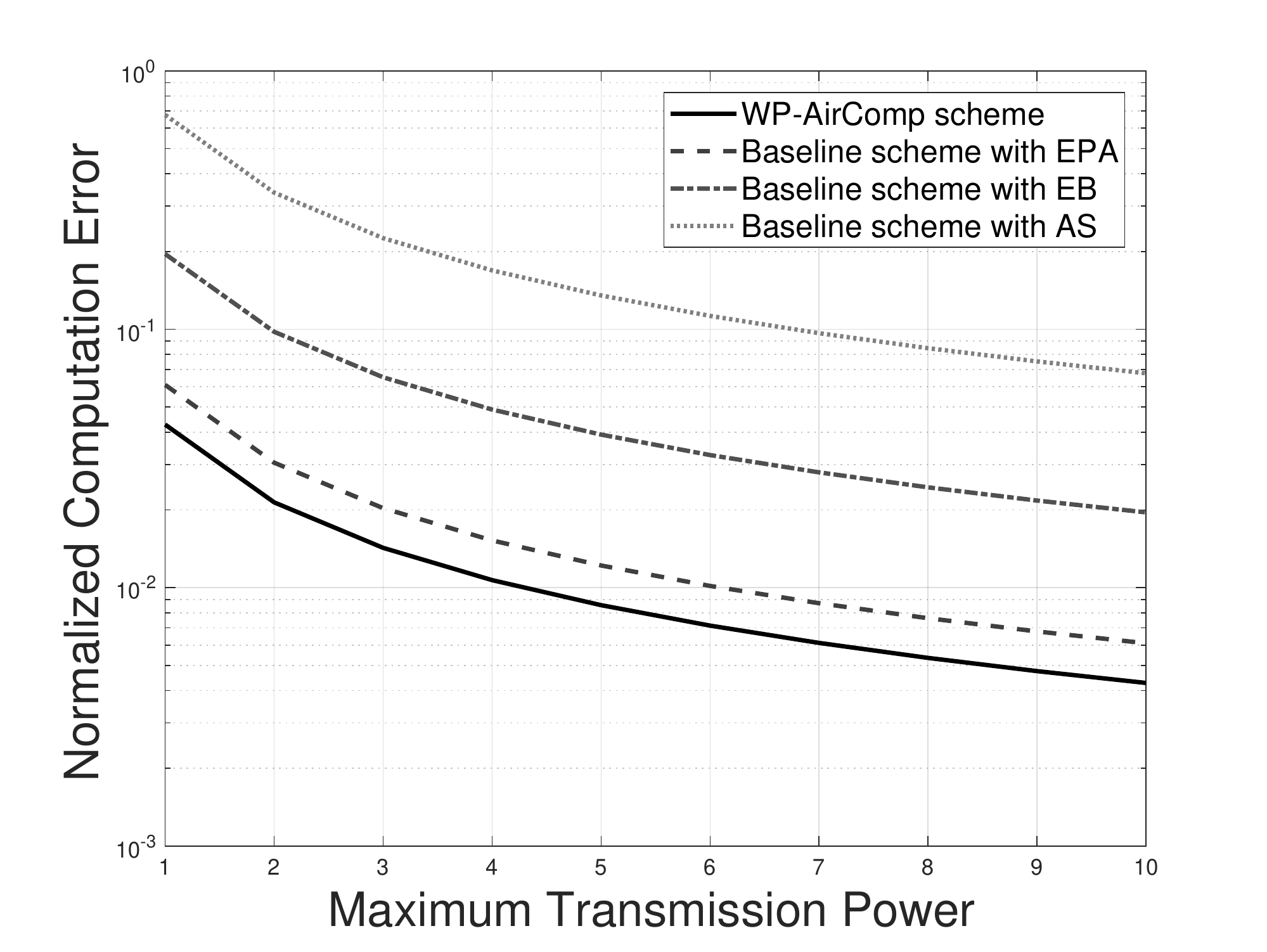}}
  \subfigure[MIMO case]{
  \label{MIMO_P}
  \includegraphics[scale=0.38]{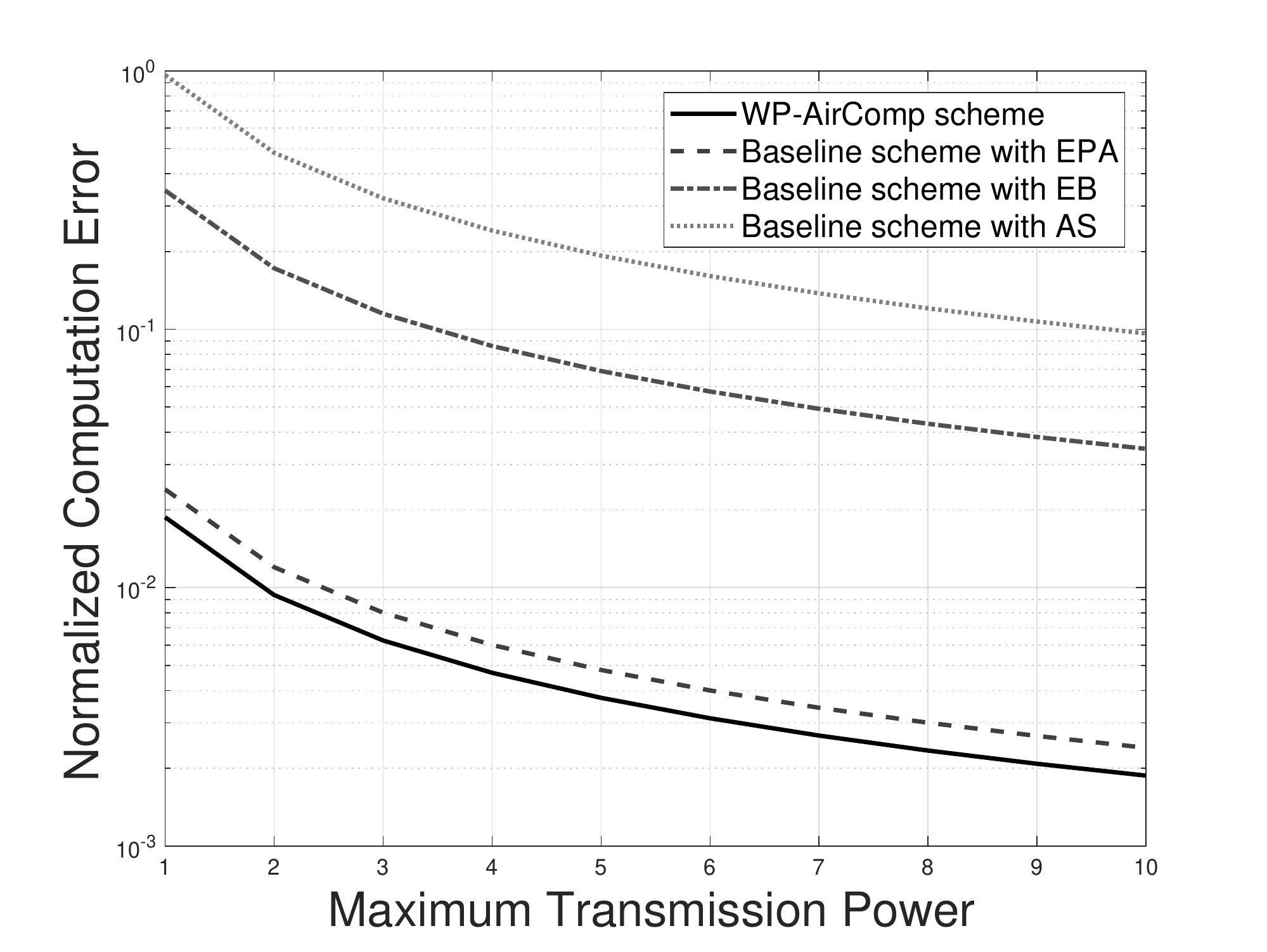}}
  \caption{The effects of server transmission power on the computation error of AirComp.}
  \label{FigP}
\end{figure}

First, the performance of the normalized computation error under varying maximum transmission power are evaluated in Fig.~\ref{FigP} for both the MISO and MIMO cases. It can be observed that the normalized computation error decreases with the increasing maximum transmission power, which coincides with the intuition that larger transmitting power will result in higher SNR and thus less estimation error. Moreover, our proposed solution outperforms all other baseline schemes throughout the whole considered range of transmission power. Particularly, there are significant gaps between the baseline schemes of AS, EB and our proposed WP-AirComp scheme, which demonstrate the merit of solving the outer sub-problem for aggregation beamformer optimization. Interestingly, the gap between the baseline scheme of EPA and the WP-AirComp scheme is relatively narrow, suggesting that optimizing the power allocation is not so critical in terms of AirComp accuracy. 

\begin{figure}[t]
  \centering
  \subfigure[MISO case]{
  \label{MISO_Nr}
  \includegraphics[scale=0.38]{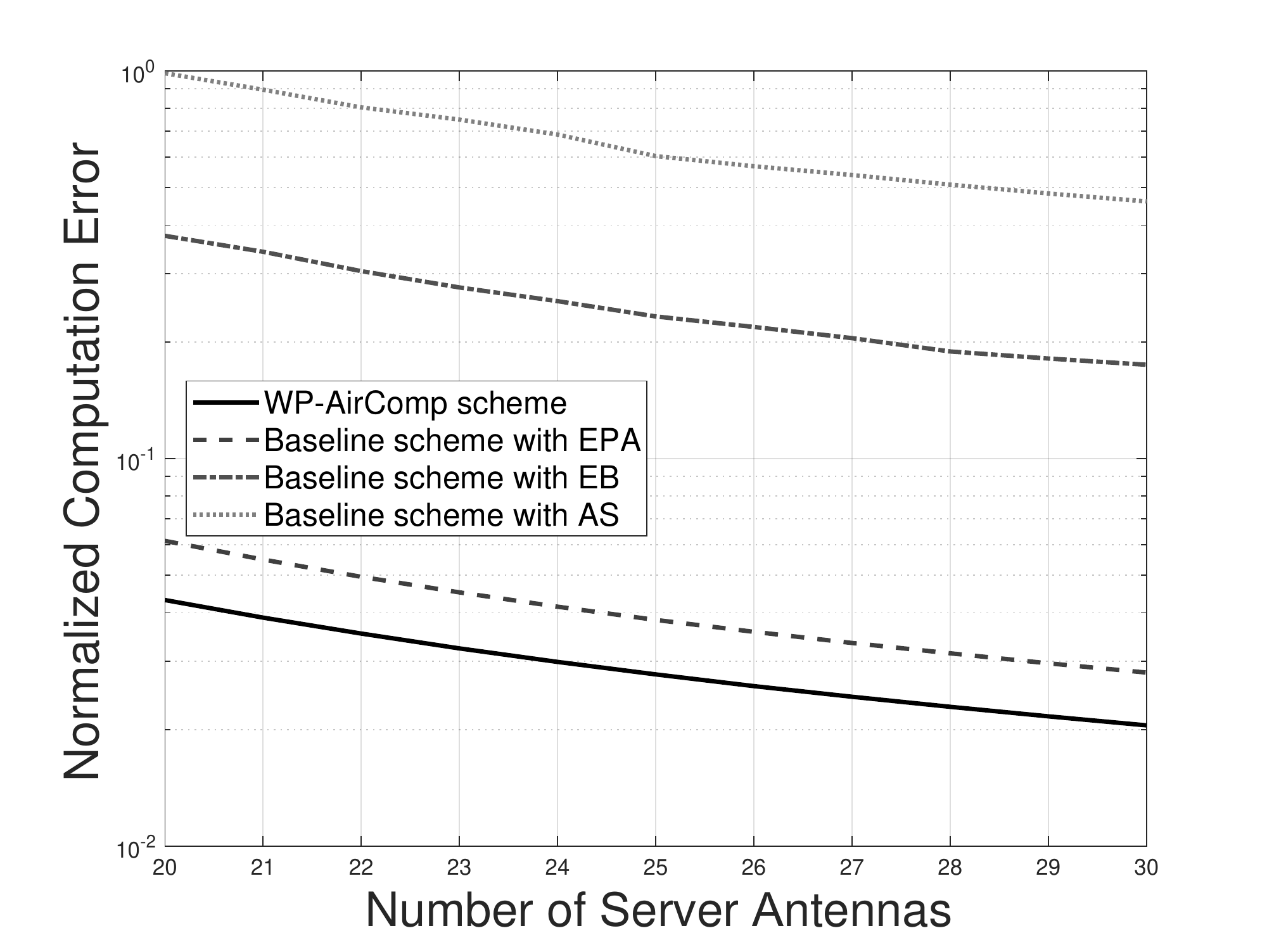}}
  \subfigure[MIMO case]{
  \label{MIMO_Nr}
  \includegraphics[scale=0.38]{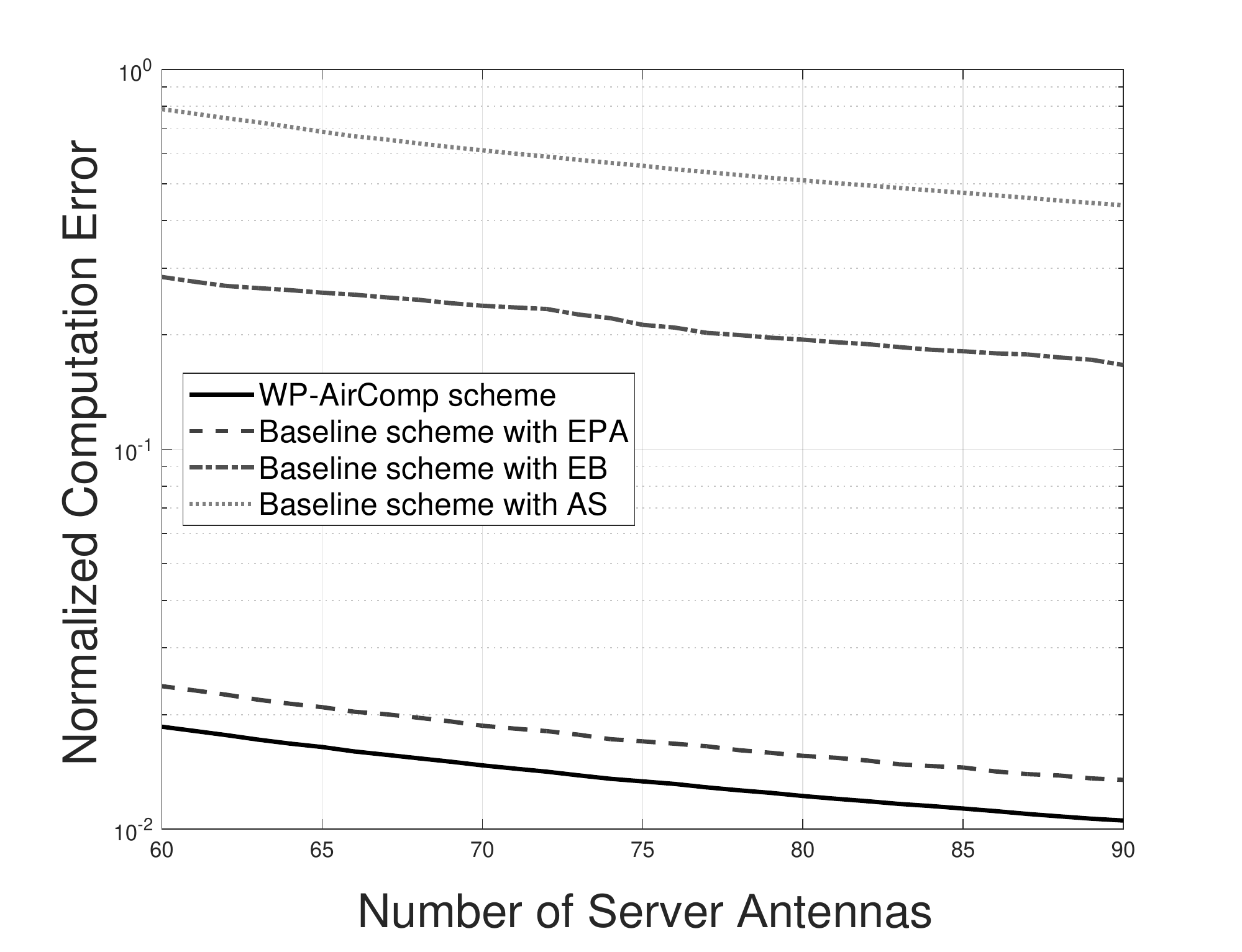}}
  \caption{The effects of server antenna numbers on the computation error of AirComp.}
  \label{FigNr}
\end{figure}

Fig.~\ref{FigNr} further compares the proposed solution with all three baseline schemes under varying number of  antennas at the server. It is observed that, for all four schemes, the normalized computation error monotonically decreases with the growth of the antenna numbers at the server due to the increasing diversity gain. Similar observations to those in Fig. \ref{FigP} also apply here, showing that our proposed WP-AirComp scheme performs consistently well in various settings of the system parameters.

\subsection{Effects of Different System Parameters on WP-AirComp Performance}
\begin{figure}[t]
  \centering
  \subfigure[MISO case]{
  \label{MISO_K}
  \includegraphics[scale=0.38]{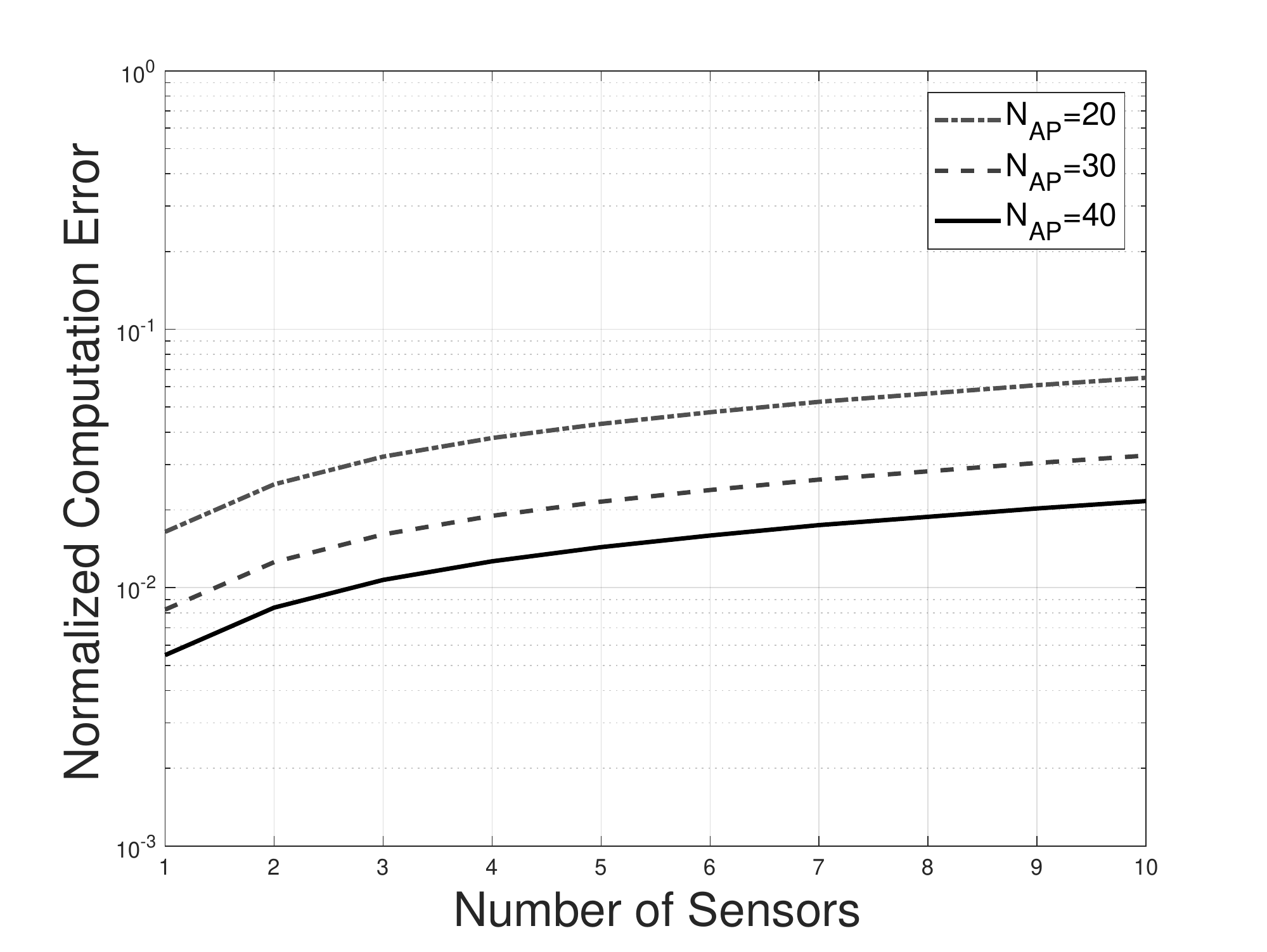}}
  \subfigure[MIMO case]{
  \label{MIMO_K}
  \includegraphics[scale=0.38]{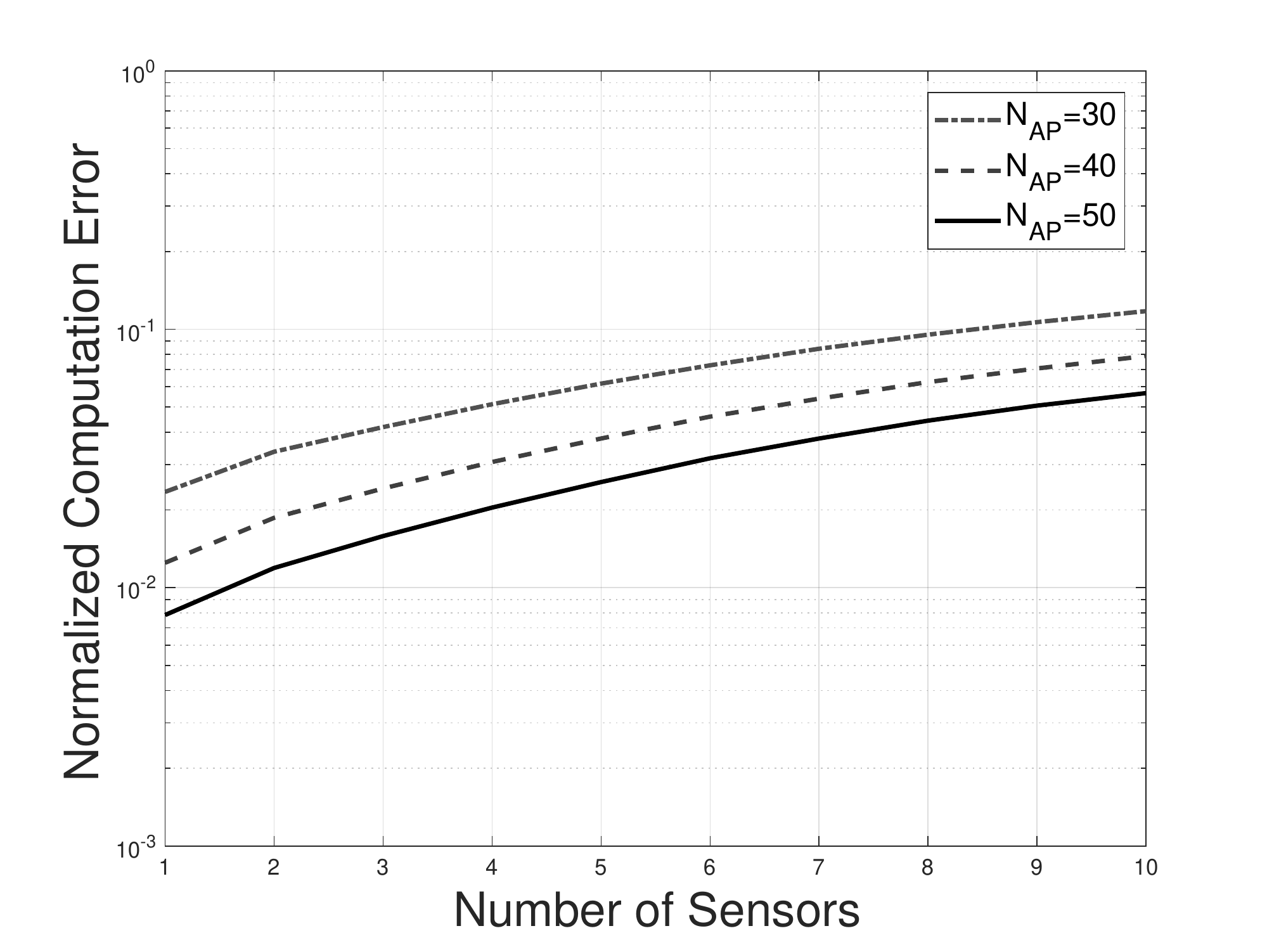}}
  \caption{The effects of sensor numbers on the computation error of AirComp.}
  \label{FigK}
\end{figure}

We further quantify the effects of different system parameters on the proposed WP-AirComp framework in Fig.~\ref{FigK}, which shows the curves of the normalized computation error versus the number of sensors under different antenna settings at the server. One can observe that the computation error increases with the number of sensors but decreases with the number of antennas at the server. This aligns with our intuition that more connected sensors makes it harder to design one common data aggregation beamformer to equalize all different sensors' channels, while having more antennas at the server can significant boost the computation accuracy by exploiting the spatial diversity gain.


\section{Concluding Remarks}
In this paper, we have proposed the WP-AirComp framework for the joint design of wireless power allocation, energy and aggregation beamforming for effective WDA in IoT systems. The framework design is tractable via an intelligent decomposition of the original non-convex problem into an outer-inner form separating the complex design into sub-problems. The decomposition approach not only yields useful insights into the optimal solution structure, but also approaches globally optimum with a high probability. The additional design dimension created by wireless power control is shown to be able to boost the AirComp accuracy. The work points to the promising new research area of WP-AirComp where many interesting research issues warrant further investigation, such as sensor scheduling, sensor clustering, and multiple servers cooperation.

\section*{Acknowledgment}
This work was supported in part by Hong Kong Research Grants Council under the Grants 17209917 and 17259416, and
by Shenzhen Science and Technology Program under Grant No. JCYJ20170817110410346.

\appendix
\subsection{Proof of Lemma \ref{MISO_ZF}}\label{App:MISO_ZF}
Given the computation-error minimization objective provided in \eqref{MSE_function}, it is easy to note that both the first and the second terms within, i.e., $\sum_{k=1}^{K} \|\bold{a}^H\bold{h}_k b_k-1\|^2$ and $\bold{a}^H\bold{a}$ are positive. As a result, for any given data precoder $\bold{a}$, we have the following inequality:
\begin{equation}\label{app:psd_ineq}
 \sum_{k=1}^{K} \|\bold{a}^H\bold{h}_k b_k-1\|^2+\sigma_n^2\bold{a}^H\bold{a} \ge \sigma_n^2 \bold{a}^H\bold{a}.
\end{equation}
It is easy to verify that setting $\{b_k\}$ to have the zero-forcing structure in \eqref{Eq:MISO_ZF} enforces  
$$\sum_{k=1}^{K} \|\bold{a}^H\bold{h}_k b_k-1\|^2 = 0,$$
and thus achieves the equality in \eqref{app:psd_ineq}, combining with $\bold{a} = \eta\bold{f}$ completes the proof.

\subsection{Proof of Lemma \ref{MISO_Equal}}\label{App:MISO_Equal}
Since $0 \le P_k \le P_0$, there exists $P_k$ such that,
$$\exists ~ C, \text{s.t.}, ~ \gamma_k \|\bold{g}_k\|^2 \|\bold{h}_k^H\bold{f}\|^2 P_k = C, ~ \forall k.$$

If $\gamma_k \|\bold{g}_k\|^2 \|\bold{h}_k^H\bold{f}\|^2 P_k$ are not equal at the optimal point. Assume that $\gamma_{\min} \|\bold{g}_{\min}\|^2 \|\bold{f}^H\bold{h}_{\min}\|^2 P_{\min}^*$ is the minimum and $\gamma_{\max} \|\bold{g}_{\max}\|^2 \|\bold{f}^H\bold{h}_{\max}\|^2 P_{\max}^*$ is the maximum. Let $P'_{\min} = P_{\min}^* + \delta$ and $P'_{\max} = P_{\max}^* - \delta$, where $\delta$ is small enough to ensure that $P'_{\min} < P'_{\max}$. Without violating the power constraint, 
$$\gamma_{\min} \|\bold{g}_{\min}\|^2 \|\bold{f}^H\bold{h}_{\min}\|^2 P'_{\min}>\gamma_{\min} \|\bold{g}_{\min}\|^2 \|\bold{f}^H\bold{h}_{\min}\|^2 P_{\min}^*.$$ 
By such contradiction, equal $\gamma_k \|\bold{g}_k\|^2 \|\bold{f}^H\bold{h}_k\|^2 P_k^*$ is proved to be the optimal strategy.

\subsection{Proof of Lemma \ref{MISO_Conv}}\label{App:MISO_Conv}
According to \cite{Boyd2006convex}, $\tr(\bold{h}_k\bold{h}_k^H\hat{\bold{F}})$ can be regarded as a linear function of $\hat{\bold{F}}$. Since the channel gain cannot be negative, i.e., $\bold{h}_k \succeq 0$, combining with the constraint $\bold{0} \preceq \hat{\bold{F}} \preceq\bold{I}$, $\tr(\bold{h}_k\bold{h}_k^H\hat{\bold{F}})$ is always positive. Based on the composition rule of scalar functions, $\frac{1}{\gamma_k \tr(\bold{h}_k\bold{h}_k^H\hat{\bold{F}})\|\bold{g}_k\|^2 P_0}$ is convex. Since the summation keeps the convexity, the objective function of Problem P10 is convex. Combining with the convex constraints leads to the result.

\subsection{Proof of Lemma \ref{MIMO_Equal}}\label{App:MIMO_Equal}
Since $0 \le P_k \le P_0$, there exists $P_k$ such that,
$$\exists ~ C, \text{s.t.}, ~ \frac{\gamma_k \sigma_{\max}^2(\bold{G}_k) P_k}{\tr\l((\bold{F}^H\bold{H}_k\bold{H}_k^H\bold{F})^{-1}\r)} = C, ~ \forall k.$$

If $\frac{\gamma_k \sigma_{\max}^2(\bold{G}_k) P_k}{\tr\l((\bold{F}^H\bold{H}_k\bold{H}_k^H\bold{F})^{-1}\r)}$ are not equal at the optimal point. Assume that $\frac{\gamma_{\min} \sigma_{\max}^2(\bold{G}_{\min}) P_{\min}^*}{\tr\l((\bold{F}^H\bold{H}_{\min}\bold{H}_{\min}^H\bold{F})^{-1}\r)}$ is the minimum and $\frac{\gamma_{\max} \sigma_{\max}^2(\bold{G}_{\max}) P_{\max}^*}{\tr\l((\bold{F}^H\bold{H}_{\max}\bold{H}_{\max}^H\bold{F})^{-1}\r)}$ is the maximum. Let $P'_{\min} = P_{\min}^* + \delta$ and $P'_{\max} = P_{\max}^* - \delta$, where $\delta$ is small enough to ensure that $P_{\min}' < P_{\max}'$. Without violating the power constraint, 
$$\frac{\gamma_{\min} \sigma_{\max}^2(\bold{G}_{\min}) P'_{\min}}{\tr\l((\bold{F}^H\bold{H}_{\min}\bold{H}_{\min}^H\bold{F})^{-1}\r)}>\frac{\gamma_{\min} \sigma_{\max}^2(\bold{G}_{\min}) P_{\min}^*}{\tr\l((\bold{F}^H\bold{H}_{\min}\bold{H}_{\min}^H\bold{F})^{-1}\r)}.$$ 
By such contradiction, equal $\frac{\gamma_k \sigma_{\max}^2(\bold{G}_k) P_k^*}{\tr\l((\bold{F}^H\bold{H}_k\bold{H}_k^H\bold{F})^{-1}\r)}$ is proved to be the optimal strategy.

\subsection{Proof of Lemma \ref{MIMO_Conv}}\label{App:MIMO_Conv}
According to \cite{Boyd2006convex}, $\lambda_{\min}(\bold{H}_k^H\hat{\bold{F}}\bold{H}_k)$ can be regarded as a concave function of $\hat{\bold{F}}$. Since the channel gain cannot be negative, i.e., $\bold{H}_k \succeq 0$, combining with the constraint $\bold{0} \preceq \hat{\bold{F}} \preceq\bold{I}$, $\lambda_{\min}(\bold{H}_k^H\hat{\bold{F}}\bold{H}_k)$ is always positive. Based on the composition rule of scalar functions, the equivalent MSE function for each sensor $\frac{L}{\gamma_k \sigma_{\max}^2(\bold{G}_k) \lambda_{\min}(\bold{H}_k^H\hat{\bold{F}}\bold{H}_k)P_0}$ is convex. Since the summation keeps the convexity, the objective function of Problem P19 is convex. Combining with the convex constraints leads to the result.




\footnotesize{

\bibliographystyle{ieeetr}

}

\end{document}